\pdfoutput=1
\documentclass[11pt,twoside,a4paper,cmspaper,final,collab]{cms-tdr}
\def\svnVersion{c54db2f}\def\svnDate{2022/07/13}\def\cmsCernNoTag{CERN-EP-2021-257}\def\cmsCernDate{\today}\def\cmsMessage{Published in the Journal of Instrumentation as \href{http://dx.doi.org/10.1088/1748-0221/17/07/P07023}{\doi{10.1088/1748-0221/17/07/P07023}.}}
\begin{document}\cmsNoteHeader{TAU-20-001}

\newcommand{\oneProngZeroPizero}{\ensuremath{\mathrm{h}^{\pm}}\xspace}
\newcommand{\oneProngOnePizero}{\ensuremath{\mathrm{h}^{\pm}\Pgpz}\xspace}
\newcommand{\oneProngTwoPizero}{\ensuremath{\mathrm{h}^{\pm}\Pgpz\Pgpz}\xspace}
\newcommand{\threeProngZeroPizero}{\ensuremath{\mathrm{h}^{\pm}\mathrm{h}^{\mp}\mathrm{h}^{\pm}}\xspace}
\newcommand{\threeProngOnePizero}{\ensuremath{\mathrm{h}^{\pm}\mathrm{h}^{\mp}\mathrm{h}^{\pm}\Pgpz}\xspace}
\newcommand{\twoProngZeroPizero}{\ensuremath{\mathrm{h}^{\pm}\mathrm{h}^{\pm/\mp}}\xspace}
\newcommand{\twoProngOnePizero}{\ensuremath{\mathrm{h}^{\pm}\mathrm{h}^{\pm/\mp}\Pgpz}\xspace}
\newcommand{\Pgtm}{\ensuremath{\Pgt^{-}}\xspace}
\newcommand{\hminus}{\ensuremath{\mathrm{h}^{-}}\xspace}
\newcommand{\hplus}{\ensuremath{\mathrm{h}^{+}}\xspace}
\newcommand{\hpm}{\ensuremath{\mathrm{h}^{\pm}}\xspace}
\newcommand{\Djet}{\ensuremath{D_\text{jet}}\xspace}
\newcommand{\De}{\ensuremath{D_\Pe}\xspace}
\newcommand{\Dm}{\ensuremath{D_\Pgm}\xspace}
\newcommand{\mvis}{\ensuremath{m_\text{vis}}\xspace}
\newcommand{\mtauh}{\ensuremath{m_{\tauh}}\xspace}
\newlength\cmsTabSkip\setlength{\cmsTabSkip}{1ex}

\cmsNoteHeader{TAU-20-001}
\title{Identification of hadronic tau lepton decays using a deep neural network}

\author*[inst1]{CMS Collaboration}

\date{\today}

\abstract{
   A new algorithm is presented to discriminate reconstructed hadronic decays of tau leptons (\tauh) that originate from genuine tau leptons in the CMS detector against \tauh candidates that originate from quark or gluon jets, electrons, or muons.    The algorithm inputs information from all reconstructed particles in the vicinity of a \tauh candidate and employs a deep neural network with convolutional layers to efficiently process the inputs.    This algorithm leads to a significantly improved performance compared with the previously used one.    For example, the efficiency for a genuine \tauh to pass the discriminator against jets increases by 10--30\% for a given efficiency for quark and gluon jets.    Furthermore, a more efficient \tauh reconstruction is introduced that incorporates additional hadronic decay modes.    The superior performance of the new algorithm to discriminate against jets, electrons, and muons and the improved \tauh reconstruction method are validated with LHC proton-proton collision data at $\sqrt{s} = 13\TeV$.
}

\hypersetup{pdfauthor={CMS Collaboration},pdftitle={Identification of hadronic tau lepton decays using a deep neural network},pdfsubject={CMS},pdfkeywords={CMS, tau lepton, reconstruction, machine learning, deep neural networks, convolutional neural networks}}

\maketitle

\section{Introduction}

Tau leptons play a crucial role in determining the nature of the 125\GeV Higgs boson~\cite{Sirunyan:2017khh,Sirunyan:2017djm,ATLAS:2018ynr,ATLAS:2018uni,ATLAS:2020evk}, in searching for additional Higgs bosons~\cite{Sirunyan:2018zut,ATLAS:2018gfm,Sirunyan:2018pzn,Sirunyan:2019xjg,Sirunyan:2019shc,CMS:2019hvr,Sirunyan:2019hkq,ATLAS:2020zms}, and in searching for other new particles~\cite{CMS:2019eln,Sirunyan:2018vhk,ATLAS:2019qpq}.
Since tau leptons most frequently decay to hadrons and neutrinos,
all these analyses require the efficient reconstruction and identification of hadronic tau lepton decays (\tauh).
In this paper, we introduce a new algorithm to identify \tauh candidates coming from hadronic tau lepton decays in the CMS detector and measure its performance with collision data from the CERN LHC.
The new algorithm is based on a deep neural network (DNN) that significantly improves the performance of the \tauh identification.
Furthermore, we discuss an optimized \tauh reconstruction algorithm that includes additional \tauh decay modes.

Hadronic tau lepton decays are reconstructed and identified as follows.
First, \tauh candidates are reconstructed in one of their decay modes with the ``hadrons-plus-strips'' algorithm~\cite{Chatrchyan:2012zz,Khachatryan:2015dfa,Sirunyan:2018pgf}, which reconstructs its decay mode and the visible four-momentum, \ie the four-momentum of all decay products except for the neutrinos.
Second, the \tauh candidates are identified as coming from genuine \tauh decays rather than from jets, electrons, and muons.
Quark and gluon jets often constitute the most important source of background since they are produced abundantly in proton-proton ($\Pp\Pp$) collisions at the LHC.

Previously, \tauh candidates were reconstructed only in their main decay modes: one-prong decays with and without $\Pgpz$s, and three-prong decays without $\Pgpz$s.
Now, we also consider three-prong \tauh decays with $\Pgpz$s, which account for 7.4\% of all \tauh decays.
Furthermore, three-prong \tauh decays are recovered in which one of the charged hadrons is either not reconstructed as a charged hadron or not even as a track.

The identification of \tauh candidates in CMS previously proceeded with separate discriminators against jets, electrons, and muons~\cite{Khachatryan:2015dfa,Sirunyan:2018pgf}.
The discriminators against jets and electrons combined information from high-level input variables (\ie derived quantities, such as the transverse momentum \pt sum of particles near the \tauh axis) using a multivariate analysis (MVA) classifier based on an ensemble of boosted decision trees.
A similar approach is used for \tauh identification by the ATLAS Collaboration~\cite{Aad:2014rga,Aad:2015unr}.
Our previous discriminator against muons was based on explicit thresholds of a few discriminating observables.
Here, we introduce a new \tauh identification algorithm, \textsc{DeepTau}, based on a DNN that simultaneously discriminates against jets, electrons, and muons.
The DNN uses a combination of high-level inputs, similar to previous algorithms, and information from all reconstructed particles in the vicinity of the \tauh candidate.
The information from all reconstructed particles near the \tauh axis is processed with convolutional layers~\cite{Lecun:9b68,Cun90handwrittendigit,Lecun98gradient-basedlearning} in pseudorapidity-azimuth ($\eta$-$\phi$) space.
Convolutional layers have been developed in the context of image recognition and are based on the premise that a fragment of the input, \eg an image, can be processed independently of its position, exploiting the translational invariance of the problem.
In a given physics analysis, the superior performance of such an algorithm leads to an increase of the signal efficiency for a given target background rate from jets, electrons, or muons misidentified as \tauh, which typically translates directly into superior sensitivity or precision of the analysis.

Similar approaches have been proposed and employed in the context of \tauh reconstruction and identification as well as in the classification of jets.
Identification algorithms for \tauh using shallow neural networks were studied at LEP~\cite{Innocente:1992gq}.
The ATLAS Collaboration developed an algorithm to identify visible \tauh decay products with a recurrent DNN~\cite{ATL-PHYS-PUB-2019-033}.
The ATLAS and CMS Collaborations also use DNNs to identify jets as coming from decays of \PQb quarks~\cite{Sirunyan:2017ezt,ATL-PHYS-PUB-2017-003,ATL-PHYS-PUB-2020-014,Bols:2020bkb}.
In particular, the \textsc{DeepJet} algorithm~\cite{Bols:2020bkb} employs multiclass classification to identify jets as coming from gluons, leptons, \PQb quarks, \PQc quarks, or light quarks.
Convolutional layers have been used in algorithms that identify large-radius jets as originating from the decays of high-\pt heavy particles, \eg in the \textsc{DeepAK8} algorithm that identifies jets with $\pt > 200\GeV$ as coming from decays of top quarks or decays of \PW, \PZ, or Higgs bosons~\cite{Sirunyan:2020lcu}.

This paper is structured as follows.
After an overview of the CMS detector and the event reconstruction in Section~\ref{sec:CMS}, the recent advances in the \tauh reconstruction are discussed in Section~\ref{sec:tau_reco}.
Section~\ref{sec:tau_id} describes the new \tauh identification algorithm based on neural networks and gives an overview of its performance with simulated events.
The performance of the algorithm with LHC collision events is then discussed in Section~\ref{sec:performance}.
The paper is summarized in Section~\ref{sec:summary}.
Tabulated results are provided in the HEPData record for this analysis~\cite{hepdata}.

\section{The CMS detector}
\label{sec:CMS}

The central feature of the CMS apparatus is a superconducting solenoid of 6\unit{m} internal diameter, providing a magnetic field of 3.8\unit{T}. 
Within the solenoid volume are a silicon pixel and strip tracker, a lead tungstate crystal electromagnetic calorimeter (ECAL), and a brass and scintillator hadron calorimeter (HCAL), each composed of a barrel and two endcap sections. Forward calorimeters extend the $\eta$ coverage provided by the barrel and endcap detectors. 
Muons are detected in gas-ionization chambers embedded in the steel flux-return yoke outside the solenoid. 
A more detailed description of the CMS detector, together with a definition of the coordinate system used and the relevant kinematic variables, is reported in Ref.~\cite{Chatrchyan:2008zzk}. 

\subsection{Event reconstruction}

The particle-flow (PF) algorithm~\cite{CMS-PRF-14-001} reconstructs and identifies each individual particle in an event, with an optimized combination of information from the various elements of the CMS detector. The energy of photons is obtained from the ECAL measurement. The energy of electrons is determined from a combination of the electron momentum at the primary interaction vertex, as determined by the tracker, the energy of the corresponding ECAL cluster, and the energy sum of all bremsstrahlung photons spatially compatible with originating from the electron track. The energy of muons is obtained from the curvature of the corresponding track. The energy of charged hadrons is determined from a combination of their momentum measured in the tracker and the matching ECAL and HCAL energy deposits, corrected for the response function of the calorimeters to hadronic showers. Finally, the energy of neutral hadrons is obtained from the corresponding corrected ECAL and HCAL energies.

For each event, hadronic jets are clustered from all the PF candidates using the infrared- and collinear-safe anti-\kt algorithm~\cite{Cacciari:2008gp, Cacciari:2011ma} with a distance parameter of 0.4. 
Jet momentum is determined as the vectorial sum of all particle momenta in the jet, and the simulation is, on average, within 5 to 10\% of the true momentum over the whole \pt spectrum and detector acceptance. Additional $\Pp\Pp$ interactions within the same or nearby bunch crossings (pileup) can contribute additional tracks and calorimetric energy depositions, increasing the apparent jet momentum. To mitigate this effect, charged particles identified as originating from pileup vertices are discarded and an offset correction is applied to correct for remaining contributions~\cite{CMS:2020ebo}. Jet energy corrections are derived from simulation to bring the measured response of jets to that of particle-level jets on average. In situ measurements of the momentum balance in  dijet, photon+jet, $\PZ$+jet, and multijet events are used to correct any residual differences in the jet energy scale between data and simulation~\cite{Khachatryan:2016kdb}. Additional selection criteria are applied to each jet to remove jets potentially dominated by instrumental effects or reconstruction failures~\cite{CMS:2020ebo}. 

The missing transverse momentum vector \ptvecmiss is computed as the negative vector sum of the transverse momenta of all the PF candidates in an event, and its magnitude is denoted as \ptmiss~\cite{Sirunyan:2019kia}. The \ptvecmiss is modified to include corrections to the energy scale of the reconstructed jets in the event. 

The candidate vertex with the largest value of summed physics-object $\pt^2$ is the primary $\Pp\Pp$ interaction vertex. The physics objects specifically used for this purpose are jets, clustered using the jet finding algorithm~\cite{Cacciari:2008gp,Cacciari:2011ma} with the tracks assigned to candidate vertices as inputs, and the associated missing transverse momentum, taken as the negative vector sum of the \pt of those jets, which include tracks from leptons.

Muons are measured in the range $\abs{\eta} < 2.4$, with detection planes made using three technologies: drift tubes, cathode strip chambers, and resistive plate chambers. The single muon trigger efficiency exceeds 90\% over the full $\eta$ range, and the efficiency to reconstruct and identify muons is greater than 96\%. Matching muons to tracks measured in the silicon tracker results in a relative \pt resolution for muons with $\pt < 100\GeV$ of 1\% in the barrel and 3\% in the endcaps. The \pt resolution in the barrel is better than 7\% for muons with $\pt < 1\TeV$~\cite{Sirunyan:2018fpa}. 

The electron momentum is estimated by combining the energy measurement in the ECAL with the momentum measurement in the tracker. The momentum resolution for electrons with $\pt \approx 45\GeV$ from $\PZ \to \Pe \Pe$ decays ranges from 1.7 to 4.5\%. It is generally better in the barrel region than in the endcaps, and also depends on the bremsstrahlung energy emitted by the electron as it traverses the material in front of the ECAL~\cite{Khachatryan:2015hwa}.

Events of interest are selected using a two-tiered trigger system. The first level, composed of custom hardware processors, uses information from the calorimeters and muon detectors to select events at a rate of around 100\unit{kHz} within a fixed latency of about 4\mus~\cite{Sirunyan:2020zal}. The second level, known as the high-level trigger, consists of a farm of processors running a version of the full event reconstruction software optimized for fast processing, and reduces the event rate to around 1\unit{kHz} before data storage~\cite{Khachatryan:2016bia}.

\subsection{Simulated event samples}

Monte Carlo simulated event samples are used to optimize the \tauh reconstruction, as input for the training of the neural network, and to validate the performance of the \tauh reconstruction and identification.
Unless explicitly mentioned, the simulated event samples include decays to all leptons (\Pell), \ie to electrons, muons, and tau leptons.
The samples listed in the following are produced separately for conditions corresponding to the 2016, 2017, and 2018 data-taking periods; a range of versions of the different generators is used as specified below.

Events from the production of $\PZ/\gamma^*$ and \PW bosons in association with jets ($\PZ$+jets and $\PW$+jets) are generated with \MGvATNLO v2.2.2 or v2.4.2~\cite{MadGraph} at leading order (LO) in perturbative quantum chromodynamics (QCD) with the MLM jet merging scheme~\cite{Alwall:2007fs}.
A separate $\PZ$+jets sample for the training, as well as diboson event samples ($\PW\PW$, $\PW\PZ$, and $\PZ\PZ$), are generated with \MGvATNLO at next-to-leading order (NLO) in perturbative QCD with the FxFx merging scheme~\cite{Frederix:2012ps}.
Events from top quark-antiquark pair (\ttbar) and single top quark production are generated with \POWHEG v2.0 at NLO in perturbative QCD~\cite{Nason:2004rx,Frixione:2007vw,Alioli:2009je,Alioli:2010xd,Re:2010bp,Campbell:2014kua}.
Events from multijet production via the strong interaction, referred to as QCD multijet production, are generated at LO using \PYTHIA 8.223, 8.226, or 8.230~\cite{Sjostrand:2014zea}.
The \PYTHIA event generator is also used to produce heavy additional gauge boson event samples at LO, $\zp\to\Pell\Pell$, with $m(\zp)$ ranging from 1 to 5\TeV.
The generation of 125\GeV Higgs boson (\PH) events via gluon fusion at NLO with \PH decays to tau leptons ($\PH\to\Pgt\Pgt$) is also performed with the \POWHEG 2.0 generator~\cite{Alioli:2008tz}.
The NNPDF 3.0 or 3.1 parton distribution functions are used as input in all the calculations~\cite{Ball:2014uwa,Ball:2017nwa}.

The \PYTHIA program is used to simulate tau lepton decays, parton showering, hadronization, and multiparton interactions, with version 8.223, 8.226, or 8.230 and tune CUETP8M1 or CP5~\cite{Sirunyan:2019dfx}.
For the $\zp\to\Pell\Pell$ samples, \TAUOLA is used instead to simulate tau lepton decays~\cite{Davidson:2010rw}.
Pileup interactions are generated with \PYTHIA and are overlaid on all simulated events according to the luminosity profile of the considered data.
All generated events are passed through a detailed simulation of the CMS detector with \GEANTfour~\cite{Agostinelli:2002hh}.
The simulated and recorded events are reconstructed with the same CMS reconstruction software.

\section{The \texorpdfstring{\tauh}{hadronic tau decay} reconstruction}
\label{sec:tau_reco}

The reconstruction of \tauh candidates is carried out with the hadrons-plus-strips algorithm.
The reconstruction proceeds in four steps, as summarized in the following and described in detail in Refs.~\cite{Chatrchyan:2012zz,Khachatryan:2015dfa,Sirunyan:2018pgf}.

First, seed regions are defined, with the goal of reconstructing one \tauh candidate per seed region.
Each seed region is defined by a reconstructed hadronic jet.
The jets for the seeding are clustered from all particles reconstructed by the PF algorithm using the anti-\kt algorithm with a distance parameter of $R = 0.4$.
All particles in an $\eta$-$\phi$ cone of radius $\Delta R \equiv \sqrt{\smash[b]{\Delta\eta\,^2 + \Delta\phi\,^2}} = 0.5$ around the jet axis are considered for the next steps of the \tauh reconstruction.

Second, \Pgpz candidates are reconstructed using ``strips'' in $\eta$-$\phi$ space in which the four-momenta of electrons and photons are added, and charged-hadron (\hpm) candidates are selected using charged particles from the PF algorithm as input.

Third, all possible \tauh candidates are reconstructed in a number of decay modes from the reconstructed charged hadrons and strips.
The \tauh four-momentum is obtained from summing the four-momenta of the charged hadrons and strips used to reconstruct the \tauh candidate in a given decay mode.
An overview of the \Pgt decay modes and their branching fractions is given in Table~\ref{tab:tau_decays}.
There are seven different reconstructed \tauh decay modes, including three new ones (with respect to the algorithm documented in Ref.~\cite{Sirunyan:2018pgf}) that target $\Pgtm\to\hminus\hplus\hminus\Pgpz\Pgngt$ decays and \tauh decays with three charged hadrons of which one is not reconstructed as a charged particle:

\begin{enumerate}
  \item \oneProngZeroPizero, targeting $\Pgtm\to\hminus\Pgngt$ decays (charge-conjugate decays are implied);
  \item \oneProngOnePizero, targeting $\Pgtm\to\hminus\Pgpz\Pgngt$ decays;
  \item \oneProngTwoPizero, targeting $\Pgtm\to\hminus\Pgpz\Pgpz\Pgngt$ decays;
  \item \threeProngZeroPizero, targeting $\Pgtm\to\hminus\hplus\hminus\Pgngt$ decays; 
  \item \threeProngOnePizero (new), targeting $\Pgtm\to\hminus\hplus\hminus\Pgpz\Pgngt$ decays;
  \item \twoProngZeroPizero (new), targeting $\Pgtm\to\hminus\hplus\hminus\Pgngt$ decays; 
  \item \twoProngOnePizero (new), targeting $\Pgtm\to\hminus\hplus\hminus\Pgpz\Pgngt$ decays.
\end{enumerate}

\begin{table}[!hbtp]
	\centering
		\topcaption
		[Tau lepton decays and their branching fractions.]
		{Decays of \Pgt leptons and their branching fractions ($\mathcal{B}$) in \%~\cite{PDG}. The known intermediate resonances of all the listed hadrons are indicated where appropriate. Charged hadrons are denoted by the symbol \oneProngZeroPizero. Although only \Pgtm decays are shown, the decays and values of the branching fractions are identical for charge-conjugate decays.}
		\begin{tabular}{llccc}
			& Decay mode & Resonance & \multicolumn{2}{c}{$\mathcal{B}$ (\%)} \\
			\hline
			\multicolumn{2}{l}{Leptonic decays} & & 35.2 & \\
			& $\Pgtm\to\Pem\Pagne\Pgngt$ & & & 17.8 \\
			& $\Pgtm\to\Pgmm\Pagngm\Pgngt$ & & & 17.4 \\ [\cmsTabSkip]
			\multicolumn{2}{l}{Hadronic decays} & & 64.8 & \\
			& $\Pgtm\to\hminus\Pgngt$ & & & 11.5 \\
			& $\Pgtm\to\hminus\Pgpz\Pgngt$ & \Pgr & & 25.9 \\
			& $\Pgtm\to\hminus\Pgpz\Pgpz\Pgngt$ & \Pai & & 9.5 \\
			& $\Pgtm\to\hminus\hplus\hminus\Pgngt$ & \Pai & & 9.8 \\
			& $\Pgtm\to\hminus\hplus\hminus\Pgpz\Pgngt$ & & & 4.8 \\
			& Other & & & 3.3 \\
		\end{tabular}
	\label{tab:tau_decays}
\end{table}

Fourth, a single \tauh candidate is chosen among all possible reconstructed \tauh candidates within a seed region.
Reconstructed \tauh candidates are subject to the following constraints:

\begin{itemize}
	\item The mass of the reconstructed \tauh candidate is required to be loosely compatible with: (i) the \Pgr resonance if reconstructed in the \oneProngOnePizero mode; (ii) the \Pai resonance if reconstructed in either of the \oneProngTwoPizero and \threeProngZeroPizero modes~\cite{Sirunyan:2018pgf}; or (iii) the expected mass spectrum including the \Pgra resonance if reconstructed in the \threeProngOnePizero modes. The mass of all \hpm candidates is assumed to be the charged-pion mass, in line with the output from the PF algorithm.
	\item The \tauh charge needs to correspond to $\pm 1$, unless the \tauh candidate is reconstructed in a mode with a missed charged hadron, in which case the charge is set to correspond to the charge of the charged hadron with higher \pt.
	\item All reconstructed \hpm and \Pgpz need to be in the signal cone, defined with radius $\Delta R = 3.0/\pt (\GeVns)$ (with $\Delta R$ limited to the range 0.05--0.1) with respect to the \tauh momentum.
\end{itemize}

Among the selected \tauh candidates, the one with the highest \pt is chosen.
This \tauh candidate can subsequently be discriminated against quark or gluon jets, electrons, and muons.

\begin{figure}[ht]
\centering
	\includegraphics[width=0.65\textwidth]{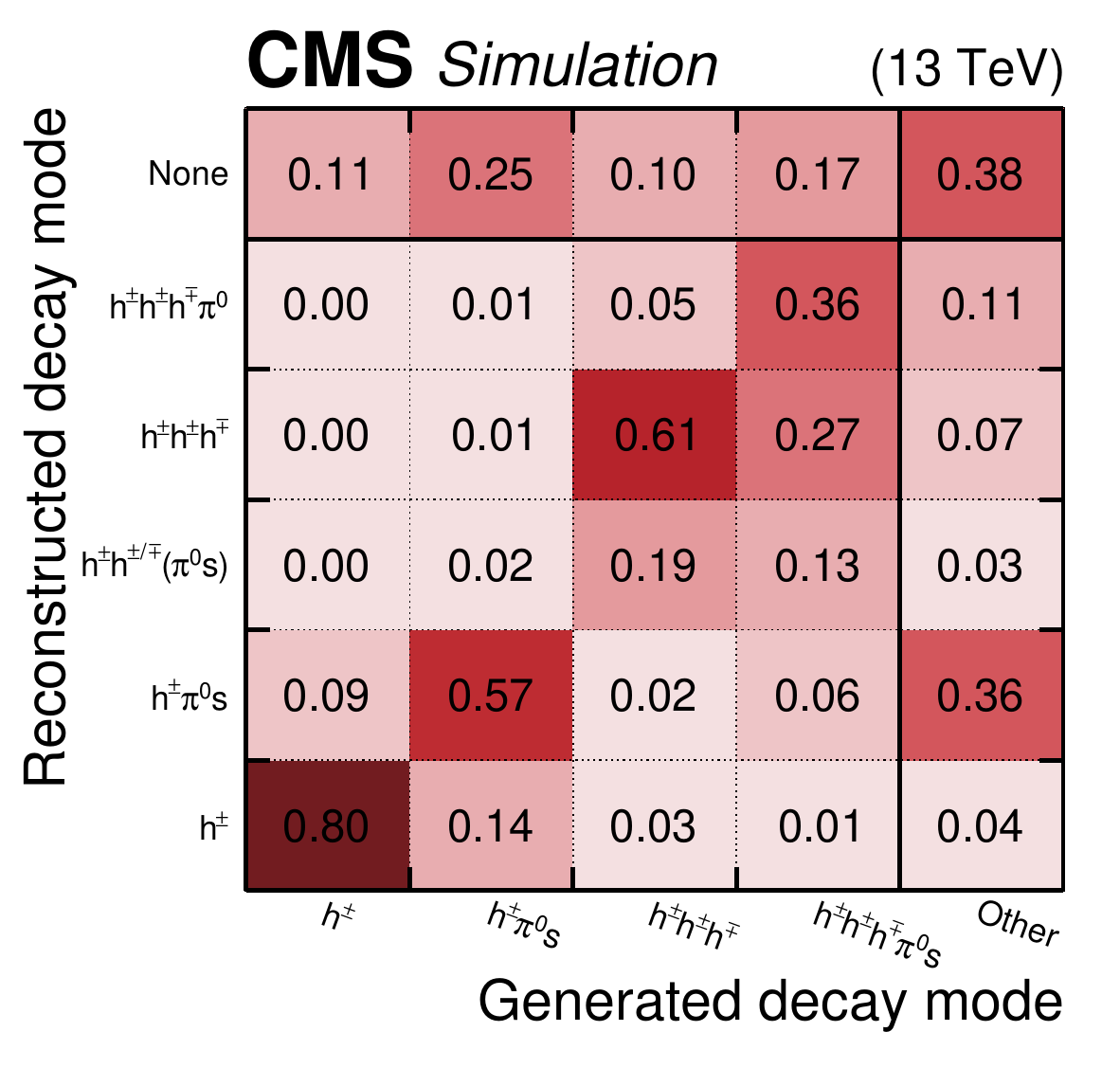}
	\caption{Decay mode confusion matrix. For a given generated decay mode, the fractions of reconstructed \tauh in different decay modes are given, as well as the fraction of generated \tauh that are not reconstructed. Both the generated and reconstructed \tauh need to fulfil $\pt > 20\GeV$ and $\abs{\eta} < 2.3$. The \tauh candidates come from a $\PZ\to\Pgt\Pgt$ event sample with $m_{\Pgt\Pgt} > 50\GeV$. Decay modes with the same numbers of charged hadrons and one or two $\Pgpz$s are combined and labelled as ``$\Pgpz$s''.}
	\label{fig:dm_migration}
\end{figure}

A large fraction of \tauh decays are reconstructed in their targeted decay modes, as inferred from Fig.~\ref{fig:dm_migration}, which shows the fractions of reconstructed decay modes for generated \tauh with a given decay mode that fulfil $\pt > 20\GeV$ and $\abs{\eta} < 2.3$.
The fraction of generated \tauh not reconstructed in any of the modes ranges from 11\% for \oneProngZeroPizero to 25\% for \oneProngOnePizero.
The overall reconstruction efficiency is mostly limited by the ability to reconstruct tracks from charged hadrons of around 90\%~\cite{CMS-PRF-14-001}.
For the decay modes without missing charged hadrons, the charge assignment is 99\% correct for an average $\PZ\to\Pgt\Pgt$ event sample, 98\% for \tauh with $\pt \approx 200\GeV$, and 92\% for \tauh with $\pt \approx 1\TeV$.

The decay modes with missing charged hadrons recover 19\% of the $\Pgtm\to\hminus\hplus\hminus\Pgngt$ decays and 13\% of the $\Pgtm\to\hminus\hplus\hminus\Pgpz\Pgngt$ decays.
However, because of the missing charged hadron, the charge assignment is only correct in
${\approx}$70\% of the cases, as opposed to an average of 99\% for the other decay modes.
Since most physics analyses apply requirements on the reconstructed \tauh charge to suppress background events, these decay modes are only useful for analyses that are not limited by background events.
None of the current analyses within the CMS Collaboration fall into this category.
Although we include the decay modes with missing charged hadrons in the new \tauh identification algorithm, most of the subsequent results are shown only for reconstructed \tauh candidates in one of the other decay modes, unless explicitly stated.

\section{The \texorpdfstring{\tauh}{hadronic tau decay} identification with a deep neural network}
\label{sec:tau_id}

The newly developed \tauh identification algorithm uses a deep neural network structure.
Its architecture is based on the following three premises: 

\begin{itemize}
	\item Multiclass: Previously, separate dedicated algorithms were used to reject electrons, muons, or quark and gluon jets reconstructed as a \tauh candidate, either based on tree ensembles (jets and electrons) or on a number of selection criteria (muons)~\cite{Sirunyan:2018pgf}.
	Including electrons, muons, and jets in the same algorithm is expected to both improve identification performance and to reduce maintenance efforts. 
	\item Usage of lower-level information: The MVA discriminators used previously were built on higher-level input variables and showed improved performance with respect to cutoff-based criteria. 
	However, jets hadronize and fragment in complex patterns, and particles from concurrent interactions lead to similarly complex detector patterns. 
	We expect that a machine-learned algorithm, which uses a sufficiently large data set for the training and is able to exploit lower-level information, can lead to improved performance. Therefore, information about all reconstructed particles near the \tauh candidate is directly used as input to the algorithm.
	\item Domain knowledge: Sets of handcrafted higher-level input variables used previously are utilized as additional inputs to the information about the single reconstructed particles. Although it should, in principle, be possible to achieve the same performance with and without using these variables given a sufficiently large set of events for the training and a suitable network architecture, the usage of the higher-level inputs may reduce the number of training events needed and improve the convergence of the training, as seen in other studies~\cite{Bols:2020bkb}.
\end{itemize}

Various network architectures were considered under the constraints that they satisfy these premises with manageable complexity. 
In particular, the algorithm must be trainable on the available hardware, and both its memory footprint and evaluation time are subject to constraints when used in the CMS reconstruction.
To process the information on all reconstructed particles near the \tauh axis, convolutional layers are employed.
Compared with other approaches like fully connected layers or graph-based structures, convolutional layers have two main advantages: they have a smaller computational complexity and they are implemented efficiently in current deep learning software.
On the other hand, they require a partitioning of inputs, in the case at hand a two-dimensional one in $\eta$-$\phi$ space.
As explained in the following, this partitioning both leads to many empty input cells and cells in which information is lost since multiple particles are found.
The details of the architecture and the inputs are given in the following.

\subsection{Inputs}

\subsubsection{Particle-level inputs}

\begin{figure}[ht]
\centering
	\includegraphics[width=0.65\textwidth]{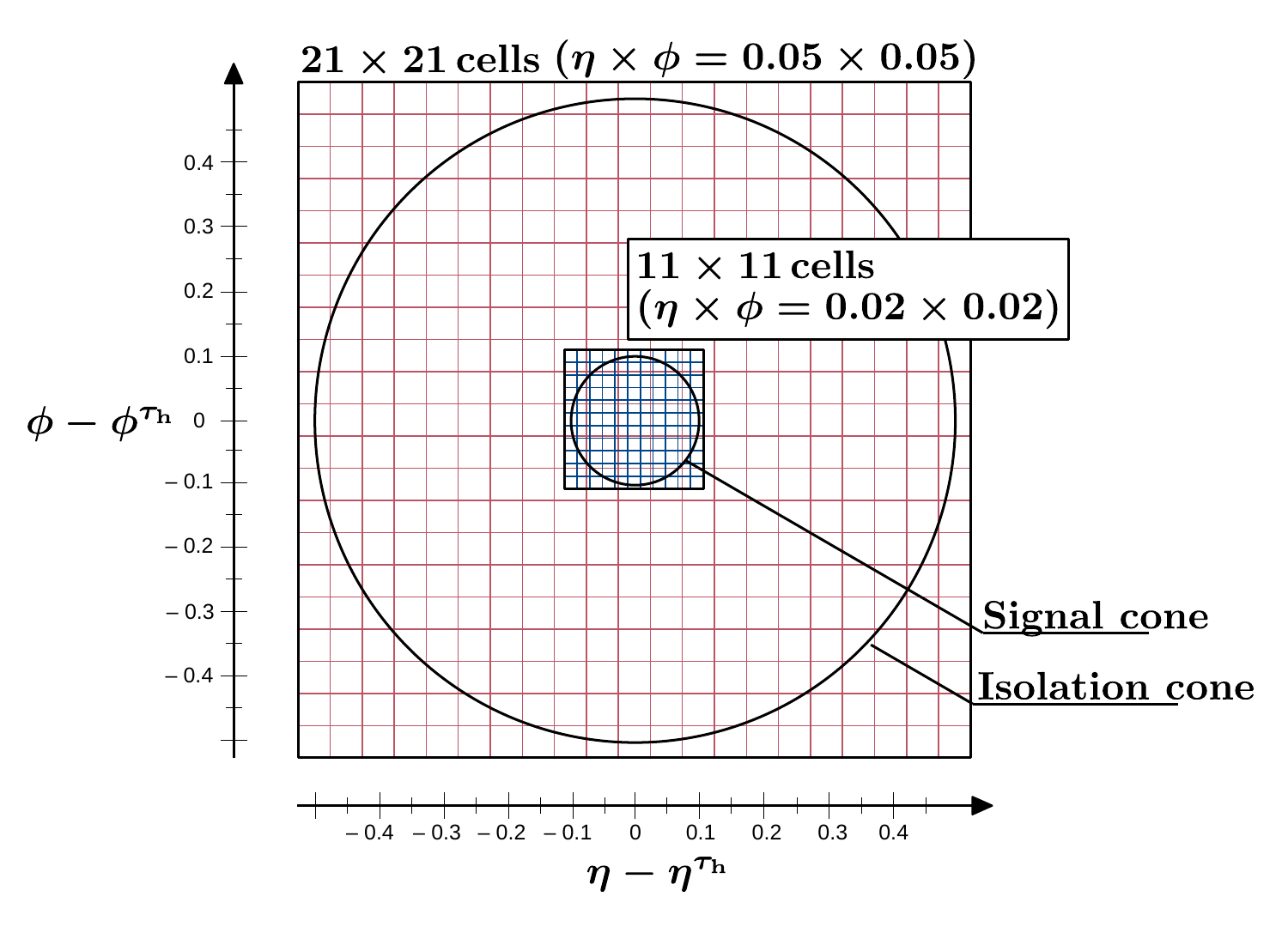}
	\caption{Layout of the grids in $\eta$-$\phi$ space around the reconstructed \tauh axis used to process the particle-level inputs for the convolutional layers of the DNN. The inner grid comprises $11{\times}11$ cells with a grid size of $0.02{\times}0.02$ and contains the signal cone with a radius of 0.05--0.1, which is defined in the \tauh reconstruction (the charged hadrons and \Pgpz candidates used to reconstruct the \tauh candidate need to be within the signal cone). For high-\pt quark and gluon jets, the finer grid is also able to resolve the dense core of the jet. The outer grid comprises $21{\times}21$ cells with a grid size of $0.05{\times}0.05$ and contains the isolation cone with a radius of 0.5 that is used to define higher-level observables that correlate with quark or gluon jet activity.}
	\label{fig:cell_definition}
\end{figure}

Two grids are defined in $\eta$-$\phi$ space, centred around the \tauh axis, as displayed in Fig.~\ref{fig:cell_definition}:
an inner grid with $11{\times}11$ cells and a grid size of $0.02{\times}0.02$, and an outer grid with $21{\times}21$ cells and grid size of $0.05{\times}0.05$.
These two grids overlap, \ie the information in the inner grid will also be part of the information that is processed by the outer grid.
For each grid cell, properties of contained reconstructed particles of seven different types are used as inputs.
If there is more than one particle of a given type, only the highest \pt particle is used as input.
The various types are the five kinds of particles reconstructed by the PF algorithm in the central detector, \ie muons, electrons, photons, charged hadrons, and neutral hadrons. The muons and electrons are reconstructed by dedicated standalone reconstruction algorithms, which provide more specific information to distinguish them from the other particles.
For each particle, various types of information are taken into account, as listed in Table~\ref{tab:inputs}.
The major tradeoffs defining the grid size are the computational cost and the loss of information when more than one particle of a specific type is present in a grid cell.
On average, 1.7\% of the inner and 7.1\% of the outer grid cells are occupied.
Up to approximately 10\% of the occupied cells contain more than one particle of a specific type.

The finer binning near the \tauh axis is motivated as follows:
First, the \tauh decay products occur predominantly in an $\eta$-$\phi$ cone with radius 0.1 around the reconstructed \tauh axis.
Second, high-\pt jets are more collimated, so a finer grid is needed to capture all particles within them.

\begin{table}[!hbtp]
	\centering
		\topcaption
		[Per-particle input variables.]
		{Input variables used for the various kinds of particles that are contained in a given cell. For each type of particle, basic kinematic quantities ($\pt, \eta, \phi$) are included but not listed below. Similarly, the reconstructed charge is included for all charged particles. An estimated per-particle probability for the particle to come from a pileup interaction using the pileup identification (PUPPI) algorithm~\cite{PUPPI} is labelled as PUPPI. A number of input variables that give the compatibility of the track with the primary interaction vertex (PV) or a possible secondary vertex (SV) from the \tauh reconstruction are denoted as ``Track PV'' and ``Track SV''.}
		\begin{tabular}{ccl}
			Particle & $N_\text{var}$ & Inputs \\
			\hline
			PF charged hadron & 27 & Track PV/SV/quality, PUPPI, HCAL energy fraction \\
			PF neutral hadron &  7 & PUPPI, HCAL energy fraction \\ [\cmsTabSkip]
			Electron 		  & 37 & Electron track quality, track/cluster matching, cluster shape \\
			PF electron		  & 22 & Track PV/SV/quality, PUPPI \\
			PF photon         & 23 & Track PV/SV/quality, PUPPI \\ [\cmsTabSkip]
			Muon			  & 37 & Track quality, muon station hits, ECAL deposits \\
			PF muon			  & 23 & Track PV/SV/quality, PUPPI \\
		\end{tabular}
	\label{tab:inputs}
\end{table}

\subsubsection{High-level inputs}

The high-level input variables correspond primarily to those proven to be useful in the previous MVA classifier~\cite{Sirunyan:2018pgf}.
These variables include: (i) the \tauh four-momentum and charge; (ii) the number of charged and neutral particles used to reconstruct the \tauh candidate; (iii) isolation variables; (iv) the compatibility of the leading \tauh track with coming from the PV; (v) the properties of a secondary vertex in case of a multiprong \tauh; (vi) observables related to the $\eta$ and $\phi$ distributions of energy reconstructed in the strips; (vii) observables related to the compatibility of the \tauh candidate with being an electron; and (viii) the estimated pileup density in the event.
In all, there are 47 high-level input variables.

\subsubsection{Transformations}

Input variables that are either integer in nature or have a significantly skewed distribution are modified with a linear transformation such that the values lie in the intervals $[-1, 1]$ or $[0, 1]$.
Most of the other input variables are subject to a standardization procedure, $x\to(x-\mu)/\sigma$, with $\mu$ denoting the mean of the distribution of $x$ and $\sigma$ its standard deviation.
The standardized inputs are truncated to $[-5, 5]$ to protect against outliers.

\subsection{Architecture}

\begin{figure}[h!]
\centering
	\includegraphics[width=\textwidth]{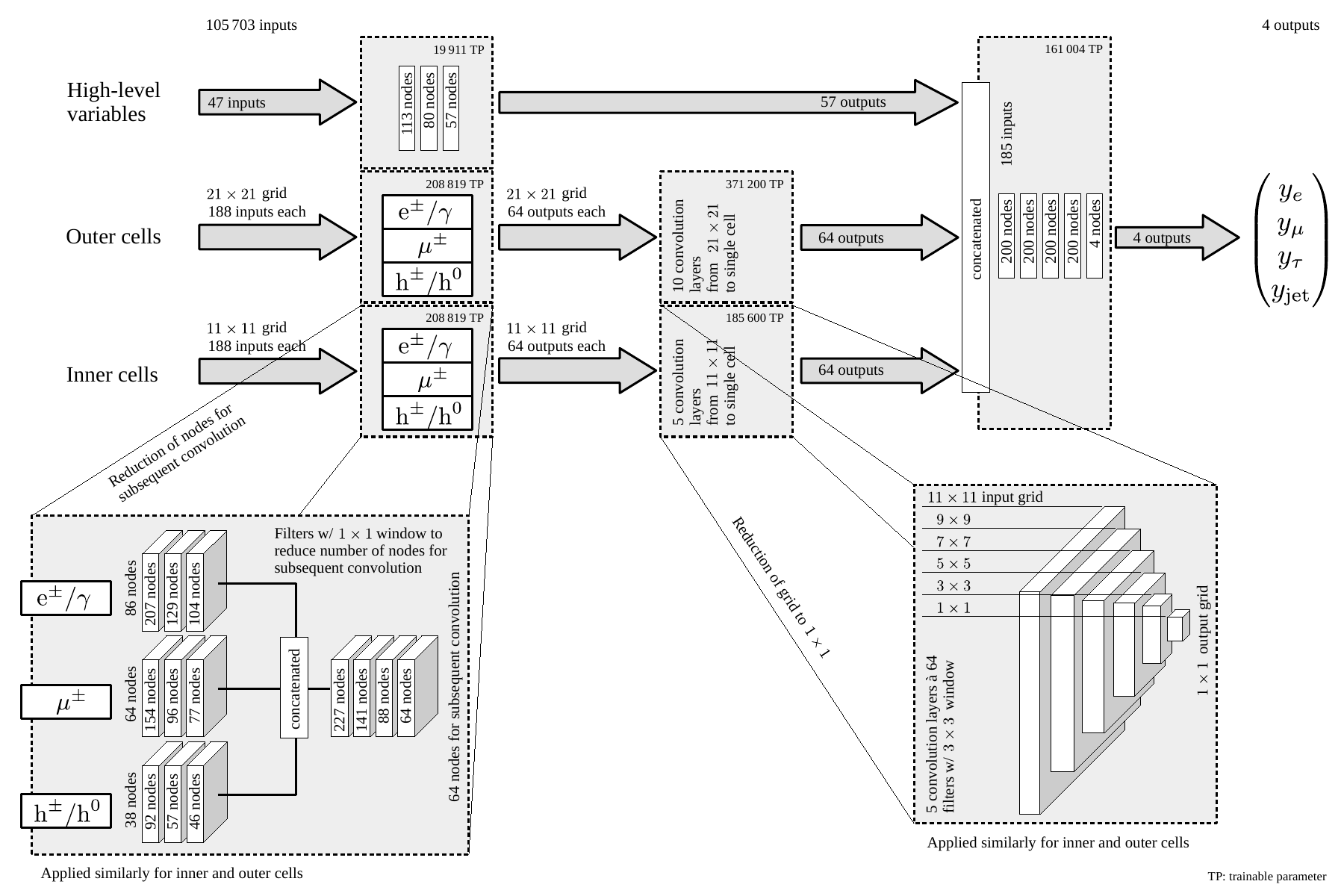}
	\caption{The DNN architecture. The three sets of input variables (inner cells, outer cells, and high-level features) are first processed separately through different subnetworks, whose outputs are then concatenated and processed through five fully connected layers before the output is calculated that gives the probabilities for a candidate to be either a \tauh, an electron, a muon, or a quark or gluon jet. The subnetwork for the high-level inputs consists of three fully connected layers with decreasing numbers of nodes, taking 47 inputs and yielding 57 outputs. The features of both the inner and outer cells are input to complex subnetworks. In the first part, the observables in each grid cell are processed through a set of fully connected layers, first separately for electrons/photons (containing both the features for PF electrons and electrons from the standalone reconstruction), muons (similarly containing both features from PF and standalone muons), and charged/neutral hadrons, passing through three fully connected layers each. The outputs are concatenated and passed through four additional fully connected layers, yielding 64 outputs for each cell. The grids are then processed with convolutional layers, which successively reduce the size of the grid. For the inner cells, there are hence 5 convolutional layers that reduce the grid from $11{\times}11$ to a single cell; for the outer cells, there are 10 convolutional layers that reduce the grid from $21{\times}21$ to a single cell. The numbers of trainable parameters (TP) for the different subnetworks are also given for the different subnetworks.}
	\label{fig:architecture}
\end{figure}

The overall DNN architecture is visualized in Fig.~\ref{fig:architecture}.
The goal of the DNN is to use and process the inputs to optimally classify the \tauh candidate as belonging to a target class, which corresponds to determining whether the reconstructed \tauh originates from a genuine \tauh, a muon, an electron, or a quark or gluon jet.
Three different subnetworks are created that independently process the inputs from the high-level variables, the outer cells, and the inner cells. 
The outputs from these three subnetworks are then concatenated and passed through four fully connected layers with 200 nodes each and then to a final layer with four nodes that yield outputs $x_\alpha$, with $\alpha\in\{\text{jet}, \Pgm, \Pe\}$.
A softmax activation function~\cite{GoodBengCour16},
\begin{linenomath}
\begin{equation}
\label{eq:y}
	y_\alpha = \exp(x_\alpha)/\Sigma_\beta \exp(x_\beta),
\end{equation}
\end{linenomath}
is then applied to yield estimates $y_\alpha$ of the probabilities for the \tauh candidate to come from each of the four target classes (\tauh, jet, \Pgm, \Pe).

The two subnetworks that process the inputs from the inner and outer cells have similar structures.
As explained above, the main idea is to process the grids with convolutional layers~\cite{Lecun:9b68,Cun90handwrittendigit,Lecun98gradient-basedlearning}.
In our case, the large number of input parameters for each cell makes it computationally too expensive to directly process the grid cells.
To reduce the dimensionality, the inputs from each cell of the inner and outer grid are first sent through a number of fully connected layers.
Since these layers act on single cells of the grids, with the full grid later on processed with convolutional layers, these fully connected layers can also be considered as one-dimensional convolutional layers, and the nodes can be considered as filters.
For each cell, inputs from the electron, muon, and hadron blocks are first processed separately in three fully connected layers with a decreasing number of nodes.
The outputs from the three separate processing blocks are then concatenated and passed through four additional fully connected layers with a decreasing number of nodes.
The 188 input parameters are reduced to 64 output parameters (filters) for each cell that are input to the convolutional layers.

The convolutional layers each make use of 64 filters with size $3{\times}3$.
These filters reduce the size of the grid with each step by two in each dimension since no padding is applied (\ie the grid is not artificially extended with additional cells beyond the dimension of the grid).
For both the inner cells with a dimension of $11{\times}11$ and the outer cells with a dimension of $21{\times}21$, the convolutional layers are applied until the grid is reduced to a single cell.
As a consequence, there are 5 convolutional layers for the inner cells and 10 for the outer cells. 
For both the inner and the outer cells, the final single cell implies that there are 64 outputs.

The 47 high-level input variables are processed by three fully connected layers with 113, 80, and 57 nodes, yielding 57 outputs.
Taken together, the three subnetworks yield 185 output parameters that are processed through a number of fully connected layers as explained above.

After each of the fully connected and convolutional layers, batch normalization~\cite{ioffe2015batch} and dropout regularization~\cite{DBLP:journals/corr/abs-1207-0580,JMLR:v15:srivastava14a} with a dropout rate of 0.2 are applied.
Furthermore, nonlinearity is introduced by applying the parametric rectified linear unit activation function, which is given by $\alpha x$ for $x < 0$ and by $x$ for $x \geq 0$, with $\alpha$ being a trainable parameter~\cite{DBLP:journals/corr/HeZR015}.

A few alternative approaches were tried. 
First, networks using only high-level inputs and an overall lower number of trainable parameters were tested.
These showed significantly inferior performance in terms of \tauh identification efficiency versus the misidentification probabilities for jets, electrons, and muons, as clarified in Figs.~\ref{fig:rocs_jets}, \ref{fig:rocs_electrons}, and~\ref{fig:rocs_muons}.
Second, alternative approaches were tried that were based on the idea of having two-dimensional grids for the particle-level inputs, but with different input cell sizes and with differently organized convolutional layers, \eg convolutional layers with padding and pooling (combining the information from a number of nearby cells).
The chosen architecture yielded the best performance among the ones tested within the predefined computing constraints.

\subsection{Loss function and classifier training}

The loss function is a sum of three different terms, given the four target classes (jet, \Pgm, \Pe, \tauh):

\begin{enumerate}
	\item a regular cross-entropy term~\cite{GoodBengCour16} for the \tauh target class,
	\item a term implementing focal loss~\cite{focal_loss} for the binary classification of \tauh against all backgrounds combined, and
	\item another focal-loss term with three components for the classification as \Pgm, \Pe, or jet, each combined with a smoothened step function that is nonzero only for \tauh candidates that are sufficiently likely to be classified as \tauh. 
\end{enumerate}

The introduction of the two focal-loss terms gives superior performance compared with using cross-entropy only.
It is motivated by two considerations.
First, analyses with \tauh candidates in the final state, like the ones mentioned in the introduction, show the highest sensitivity for efficiencies in the range 50--80\%, whereas performance for the highest efficiencies and purities is less important.
Second, the classification performance as \tauh is more important than the classification as \Pe, \Pgm, or jet; particularly in the high-purity regime where it is less important to distinguish between the three other classes.
The full definition of the loss function is given in Appendix~\ref{app:loss}.

For the training, events from the following simulated processes are used: $\PZ$+jets (NLO), $\PW$+jets, \ttbar, $\zp\to\Pgt\Pgt$, $\zp\to\Pe\Pe$, $\zp\to\Pgm\Pgm$ (with $m(\zp)$ ranging from 1 to 5\TeV), and QCD multijet production.
For testing, additional event samples are used, including $\PH\to\Pgt\Pgt$ and $\PZ$+jets (LO) event samples, and more event samples corresponding to the processes used for the training.
The event samples are simulated and reconstructed according to the 2017 data-taking conditions. 
The prospective \tauh candidates need to fulfil $\pt > 20\GeV$ and $\abs{\eta} < 2.3$.
The \tauh candidates are sampled from the various event samples such that the contributions of the different classes (jet, \Pgm, \Pe, \tauh) and in different $(\pt, \eta)$ bins are the same.
Additional weights are added to make the distributions from the different classes uniform within each $(\pt, \eta)$ bin.
In total, around 140 million \tauh candidates are used for the training, whereas 10 million are assigned as the validation samples.

The loss function is minimized with Nesterov-accelerated adaptive momentum estimation~\cite{NAdam}, which combines the \emph{Adam} algorithm~\cite{Adam} with Nesterov momentum.
The training setup uses the \textsc{Keras} library~\cite{chollet2015keras} with \textsc{TensorFlow}~\cite{tensorflow2015-whitepaper} as backend.
The training was carried out with a single Nvidia GeForce RTX\textsuperscript{\texttrademark} 2080 graphics processing unit.
The training of the final classifier was run for 10 epochs, with the training for an epoch lasting 69 hours on average.
The best performance on the validation samples has been obtained after 7 epochs.

The final discriminators against jets, muons, and electrons are given by 
\begin{linenomath}
\begin{equation}
	D_\alpha (\boldsymbol{y}) = \frac{y_\Pgt}{y_\Pgt + y_\alpha},
\end{equation}
\end{linenomath}
with $y_\alpha$ taken from Eq.~(\ref{eq:y}) and $\alpha\in\{\text{jet}, \Pgm, \Pe\}$.
In the following, these discriminators are labelled as \textsc{DeepTau} discriminators against jets (\Djet), muons (\Dm), and electrons (\De).

\subsection{Expected performance}

The performance of the \textsc{DeepTau} discriminators is evaluated with the validation samples. 
Working points are defined to guide usage in physics analyses and to derive suitable data-to-simulation corrections.
The target \tauh identification efficiencies for the various working points used in the wide range of physics analyses with tau leptons are presented in Table~\ref{tab:wp-definition}.
The target \tauh identification efficiencies are defined as the efficiency for genuine \tauh in the $\PH\to\Pgt\Pgt$ event sample that are reconstructed as \tauh candidates with $30 < \pt < 70\GeV$ to pass the given discriminator.
The target \tauh identification efficiencies range from 40 to 98\% for \Djet, from 99.5 to 99.95\% for \Dm, and from 60 to 99.5\% for \De.

\begin{table}[htb]
	\topcaption{Target \tauh identification efficiencies for the different working points defined for the three different discriminators. The target efficiencies are evaluated with the $\PH\to\Pgt\Pgt$ event sample for \tauh with $\pt \in [30, 70]\GeV$.}
    \centering
	\begin{tabular}{lcccccccc}
		& VVTight & VTight & Tight & Medium & Loose & VLoose & VVLoose & VVVLoose \\
		\hline
		$D_\Pe$ & 60\% & 70\% & 80\% & 90\% & 95\% & 98\% & 99\% & 99.5\% \\
		$D_\Pgm$ & \NA & \NA & 99.5\% & 99.8\% & 99.9\% & 99.95\% & \NA & \NA \\
		\Djet & 40\% & 50\% & 60\% & 70\% & 80\% & 90\% & 95\% & 98\% \\
	\end{tabular}
    \label{tab:wp-definition}
\end{table}

\begin{figure}[h!tb]
	\centering
	\includegraphics[width=0.48\textwidth]{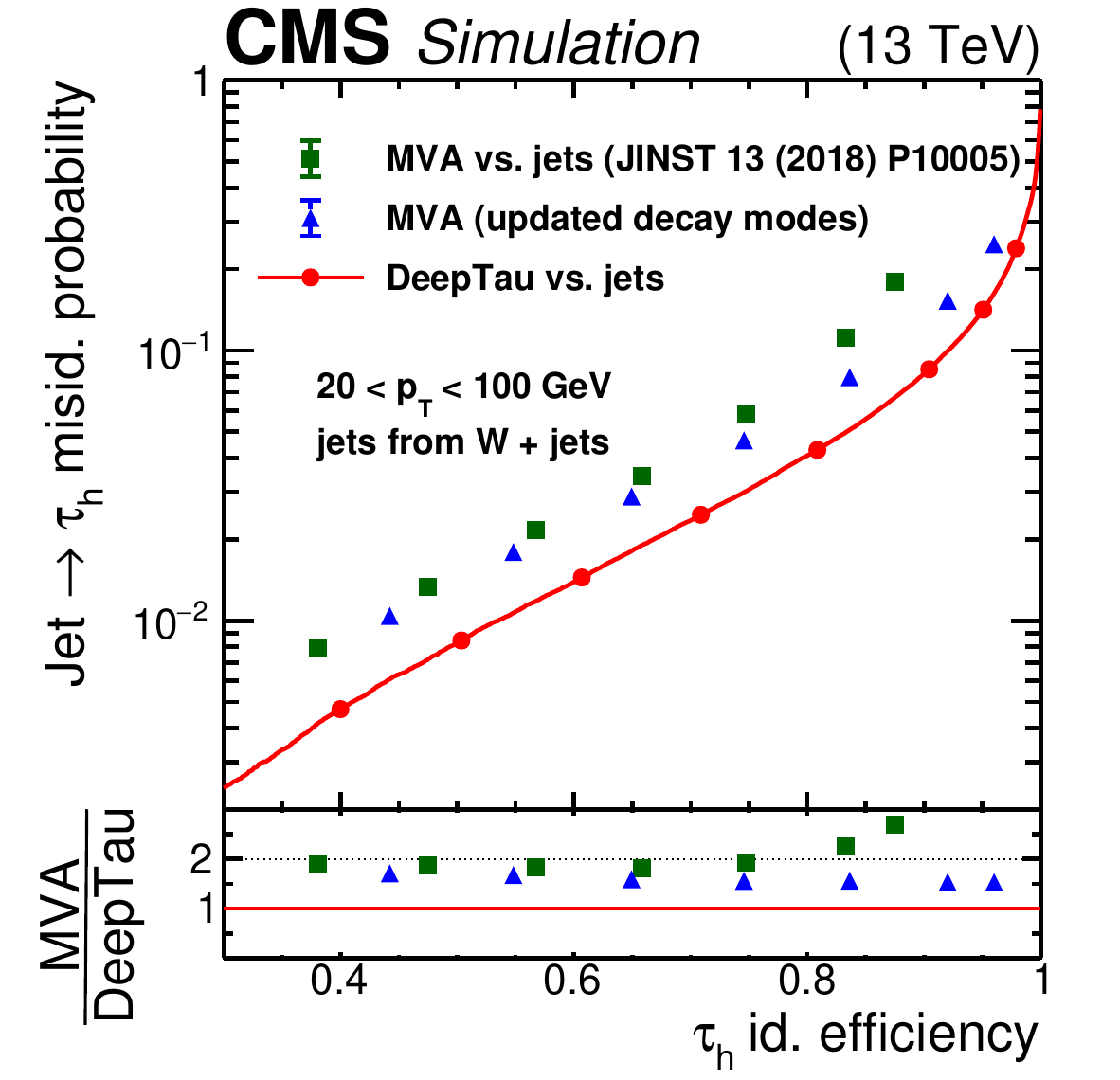}
	\includegraphics[width=0.48\textwidth]{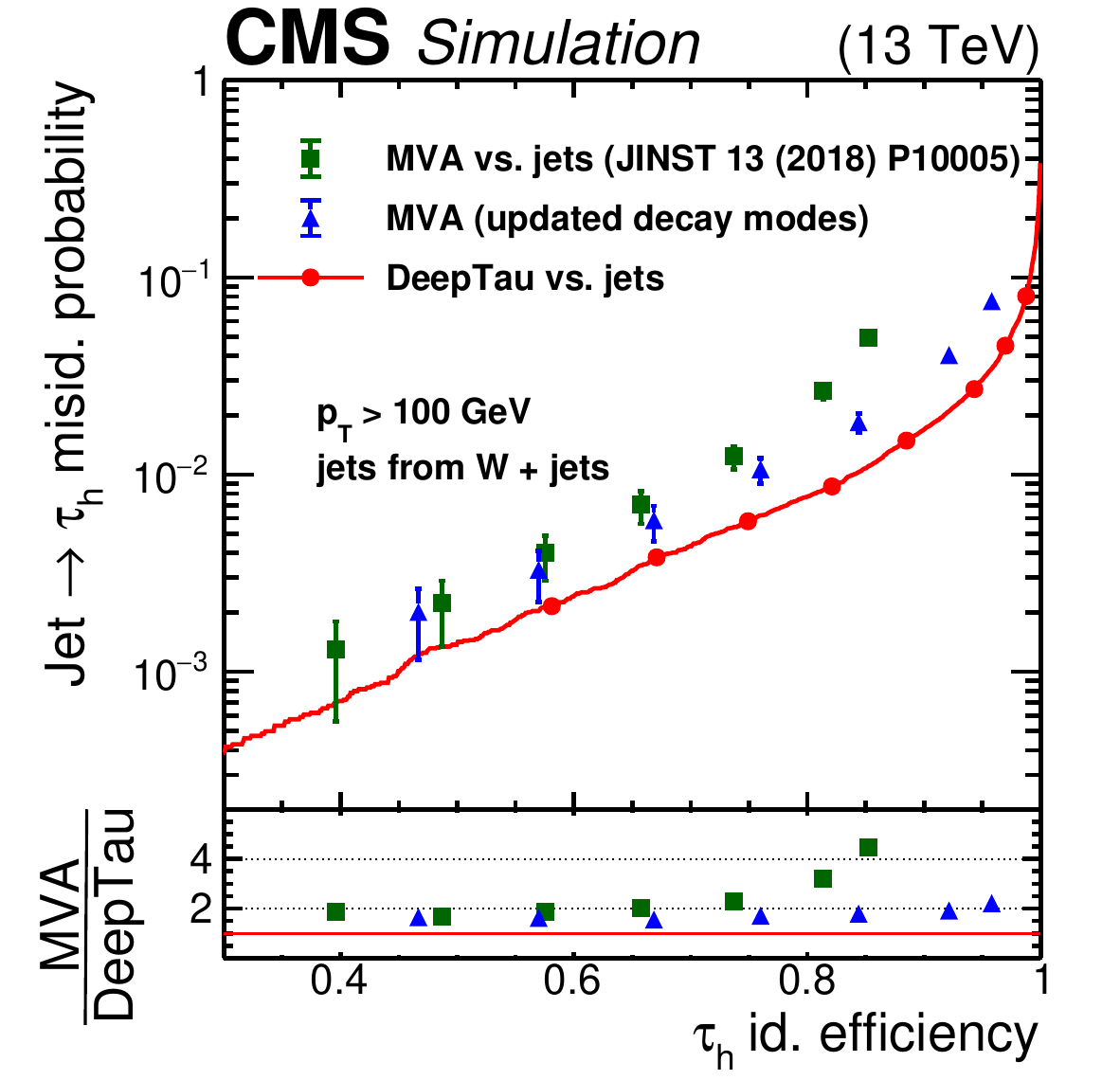}\\
	\includegraphics[width=0.48\textwidth]{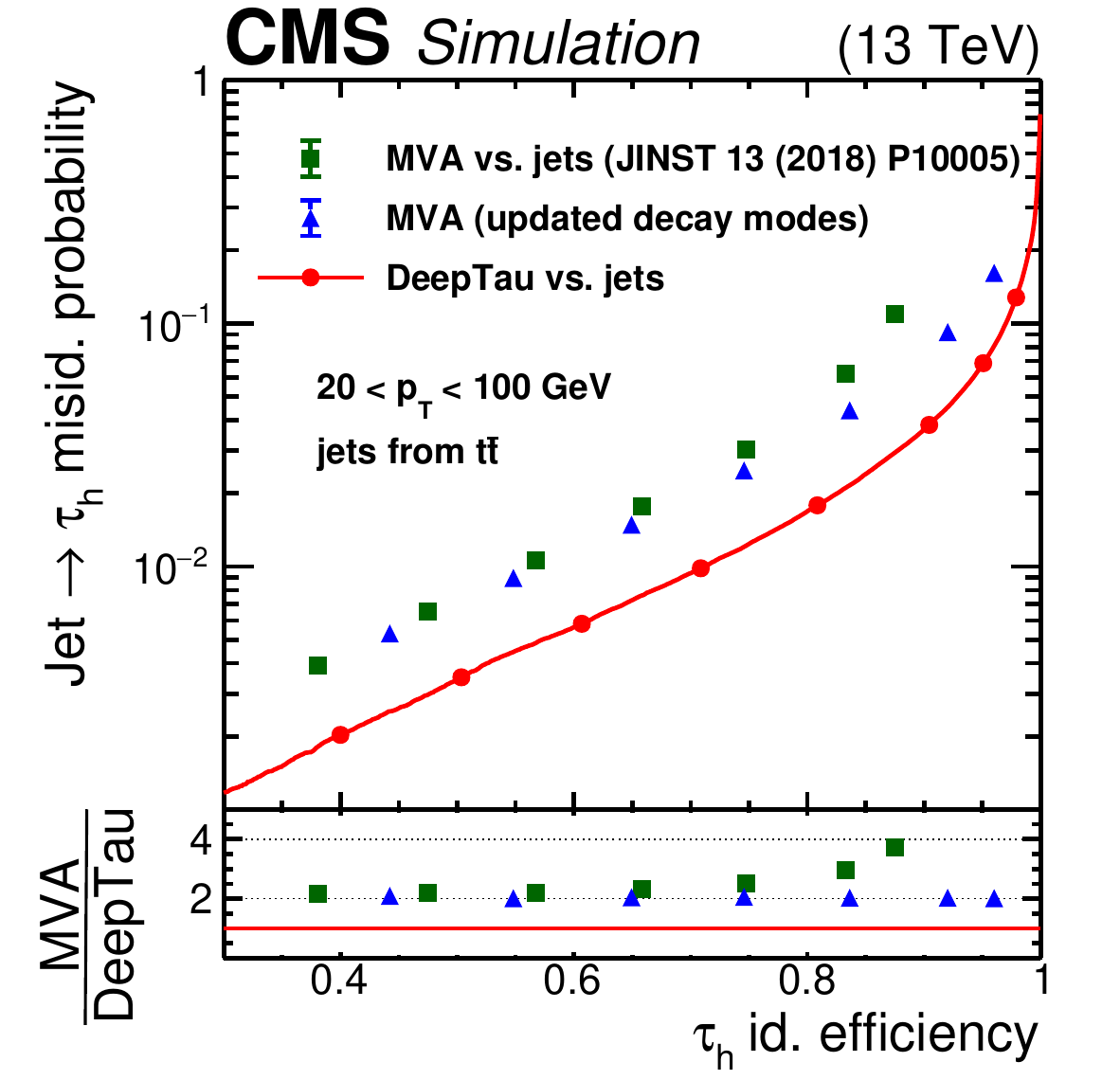}
	\includegraphics[width=0.48\textwidth]{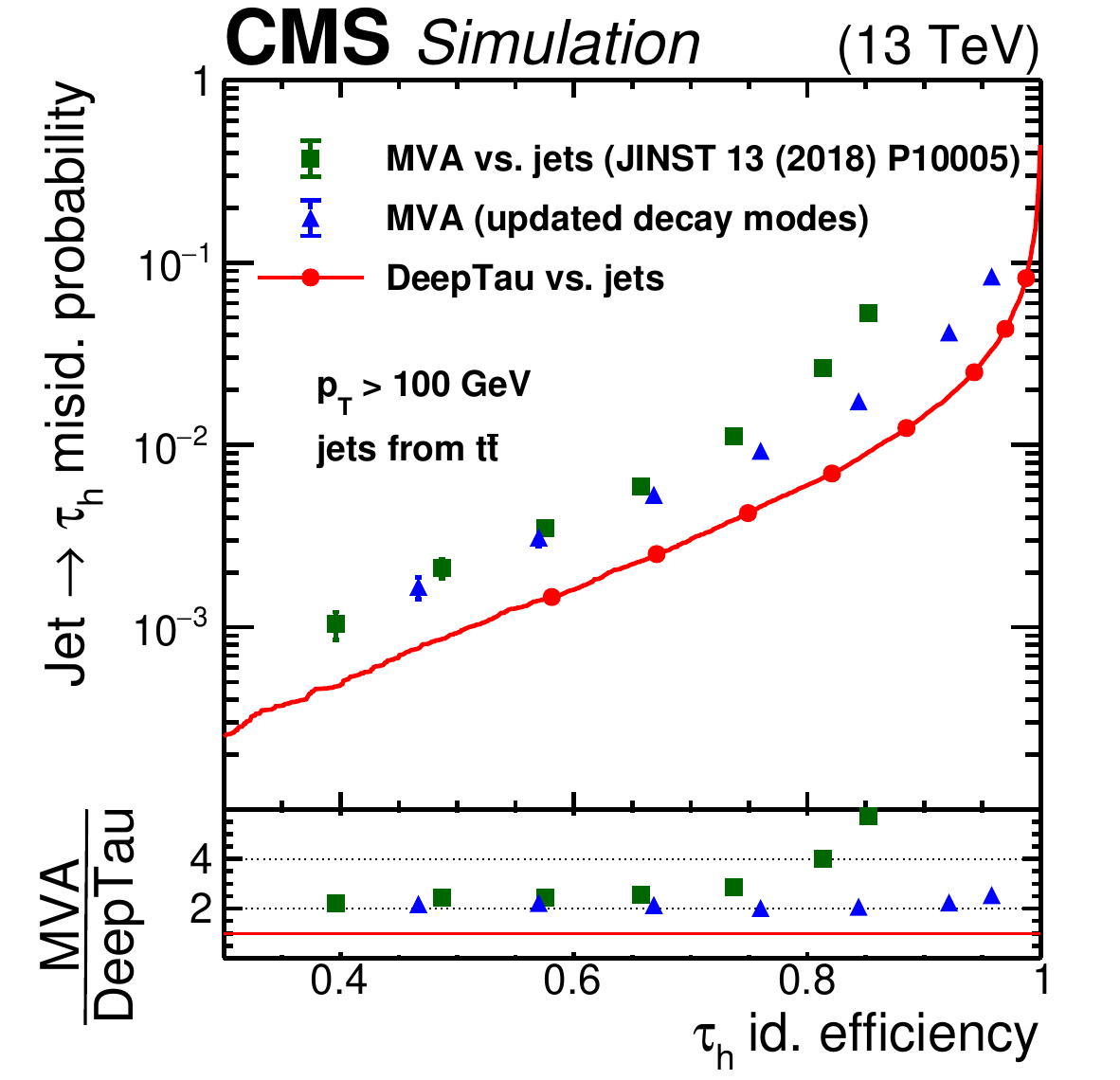}
	\caption{Efficiency for quark and gluon jets to pass different tau identification discriminators versus the efficiency for genuine \tauh. The upper two plots are obtained with jets from the $\PW$+jets simulated sample and the lower two plots with jets from the \ttbar sample. The left two plots include jets and genuine \tauh with $\pt < 100\GeV$, whereas the right two plots include those with $\pt > 100\GeV$. The working points are indicated as full circles. The efficiency for jets from the $\PW$+jets event sample, enriched in quark jets, to pass the discriminators is higher compared with jets from the \ttbar event sample, which has a larger fraction of gluon and \PQb-quark jets. The jet efficiency for a given \tauh efficiency is larger for jets and \tauh with $\pt < 100\GeV$ than for those with $\pt > 100\GeV$. Compared with the previously used MVA discriminator, the \textsc{DeepTau} discriminator reduces the jet efficiency for a given \tauh efficiency by consistently more than a factor of 1.8, and by more at high \tauh efficiency. The additional gain at high \pt comes from the inclusion of updated decay modes in the \tauh reconstruction, as illustrated by the curves for the previously used MVA discriminator but including reconstructed \tauh candidates with additional decay modes.}
	\label{fig:rocs_jets}
\end{figure}

In Fig.~\ref{fig:rocs_jets}, the performance of the \Djet discriminator is evaluated by studying the efficiencies for quark and gluon jets and genuine \tauh to pass the discriminator and comparing it with the previously used MVA classifier.
All generated and reconstructed \tauh candidates in this figure and those that follow are required to pass $\pt > 20\GeV$ and $\abs{\eta} < 2.3$, VVVLoose \De WP and VLoose \Dm WP.
For the previously used MVA classifier, two curves are shown: The first curve corresponds to the \tauh decay mode reconstruction as used in the previous publication and in previous physics analyses, whereas the second curve includes the additional \tauh decay modes introduced in Section~\ref{sec:tau_reco}, leading to an overall increase of the MVA classifier performance, in particular for higher efficiency values.
The efficiency is evaluated separately for the $\PW$+jets and \ttbar event samples.
The $\PW$+jets sample is enriched in quark jets, whereas the \ttbar event sample has a larger fraction of \PQb quark and gluon jets and has a busier event topology.
The efficiencies are also evaluated separately for \tauh candidates with $\pt < 100\GeV$ and $\pt > 100\GeV$ to test the performance in different \pt regimes.
For all considered scenarios, the \Djet discriminator shows a large improvement of the signal efficiency at a given background efficiency compared with the previously used MVA classifier.
At a given efficiency for jets to pass the discriminator, there is a relative increase in signal efficiency of more than 30\% for the tightest working points and more than 10\% for the loosest working points.
The larger gain at high \pt is a consequence of the inclusion of additional decay modes in the \tauh reconstruction.

\begin{figure}[h!]
	\centering
	\includegraphics[width=0.65\textwidth]{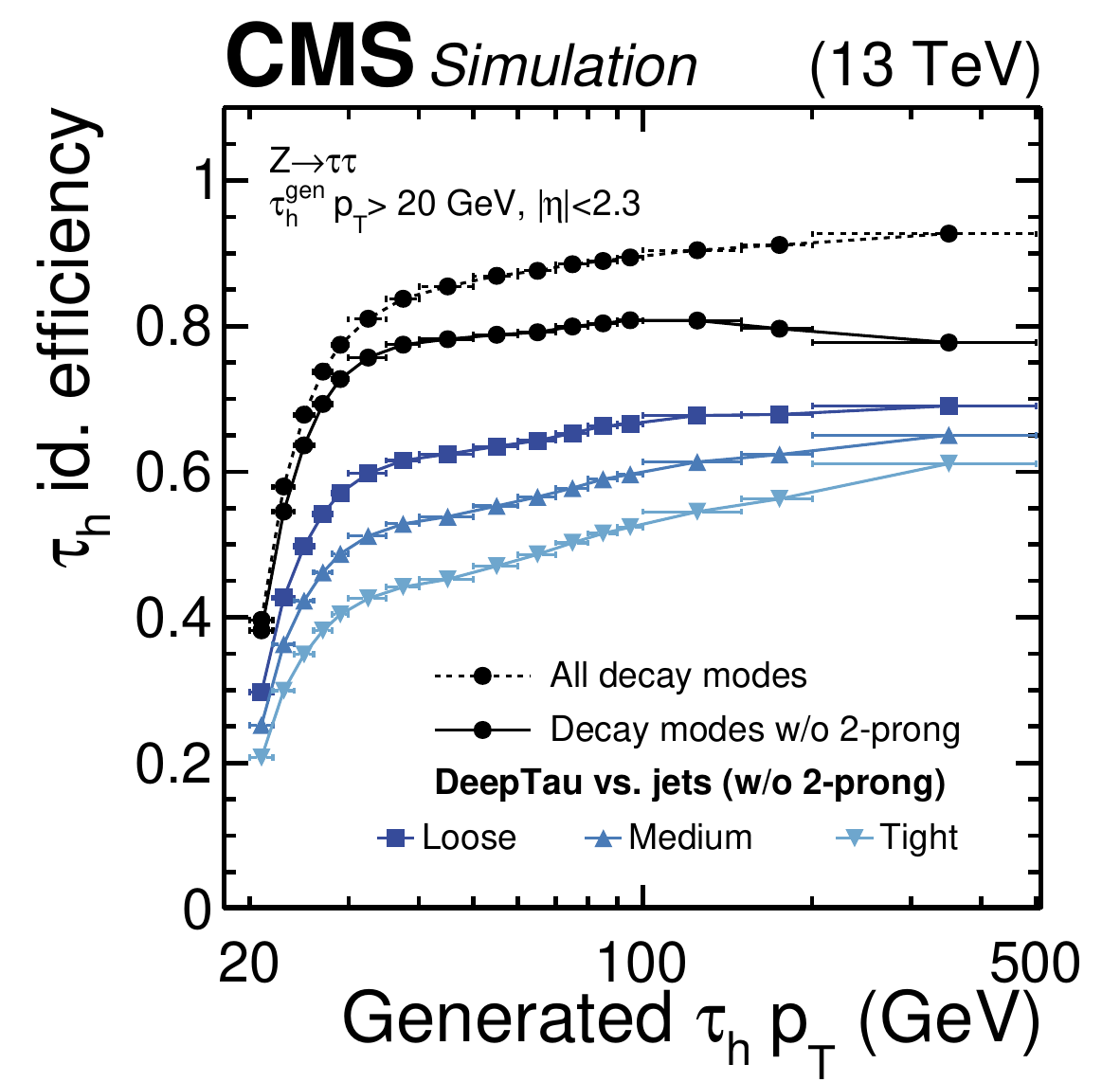}
	\caption{Efficiencies for simulated \tauh decays with $\abs{\eta} < 2.3$ to pass the following reconstruction and identification requirements: to be reconstructed in any decay mode with $\pt > 20\GeV$ and $\abs{\eta} < 2.3$ (black dashed line), to be reconstructed in a decay mode except for those with missing charged hadrons (labelled ``2-prong'' and shown as full black line), and to be reconstructed in a decay mode except the 2-prong ones and to pass the Loose, Medium, or Tight working point of the \Djet discriminator (blue lines), obtained with a $\PZ\to\Pgt\Pgt$ event sample. The efficiencies are shown as a function of the visible genuine \tauh \pt obtained from simulated decay products.}
	\label{fig:deeptau_eff_vs_pt}
\end{figure}

The \pt dependence of the efficiency for genuine generated \tauh with $\pt > 20\GeV$ to be reconstructed with $\pt > 20\GeV$ and to pass the \Djet discriminator is shown in Fig.~\ref{fig:deeptau_eff_vs_pt}.
The reconstruction efficiency exceeds 80\% for $\pt > 30\GeV$ and is close to 90\% for $\pt > 100\GeV$, and is mostly limited by the charged-hadron reconstruction efficiency.
If decay modes with missing charged hadrons (the so-called two-prong decay modes) are excluded, the efficiency is reduced by around 10\%.
The efficiency to additionally pass either of the shown working points of the \Djet discriminator (for reconstructed \tauh candidates without missing charged hadrons) is in line with the target \tauh identification efficiencies listed in Table~\ref{tab:wp-definition}.

\begin{figure}[ht]
	\centering
	\includegraphics[width=0.48\textwidth]{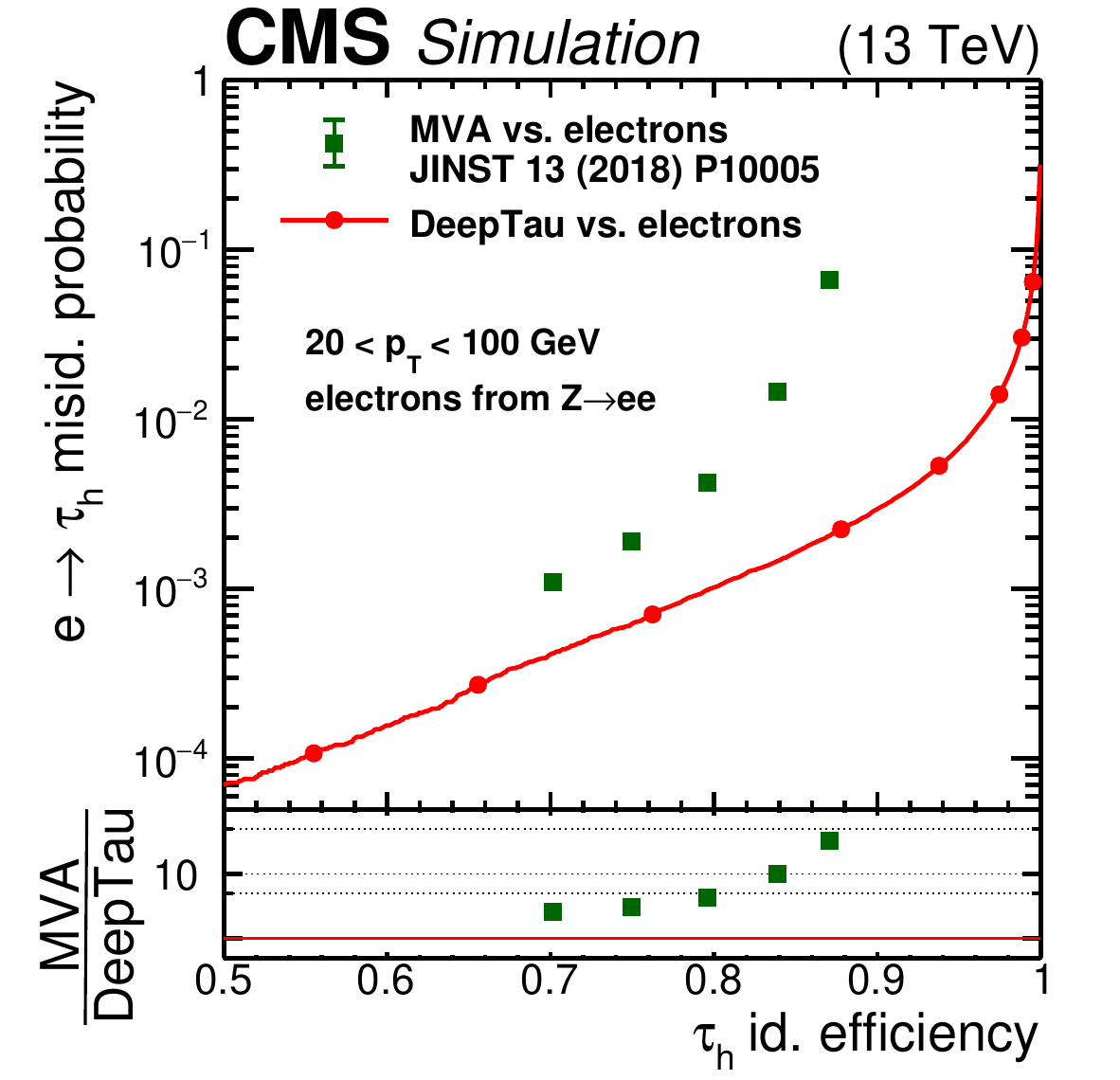}
	\includegraphics[width=0.48\textwidth]{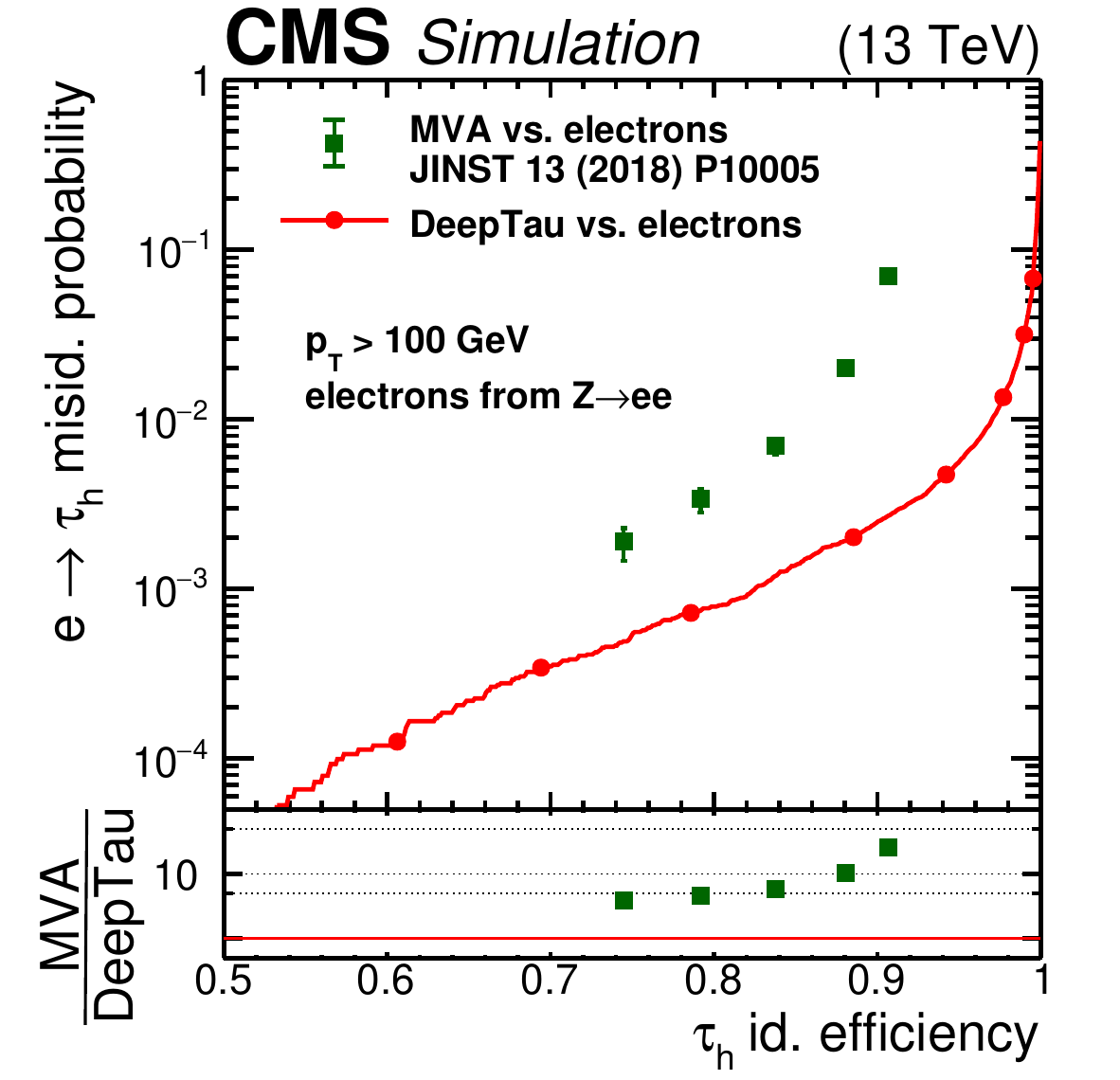}
	\caption{Efficiency for electrons versus efficiency for genuine \tauh to pass the MVA and \De discriminators, separately for electrons and \tauh with $20 < \pt < 100\GeV$ (left) and $\pt > 100\GeV$ (right). Vertical bars correspond to the statistical uncertainties. The \tauh candidates are reconstructed in one of the \tauh decay modes without missing charged hadrons. Compared with the MVA discriminator, the \De discriminator reduces the electron efficiency by more than a factor of two for a \tauh efficiency of 70\% and by more than a factor of 10 for \tauh efficiencies larger than 88\%. Furthermore, working points (indicated as full circles) are now provided for previously inaccessible \tauh efficiencies larger than 90\%, for a misidentification efficiency between 0.3 and 8\%.}
	\label{fig:rocs_electrons}
\end{figure}

The \De discriminator leads to a significantly improved rejection of electrons compared with the MVA discriminator, as can be inferred from Fig.~\ref{fig:rocs_electrons}.
The efficiencies for \tauh candidates (with $\pt > 20\GeV$, $\abs{\eta} < 2.3$ and passing the VVVLoose \Djet and VLoose \Dm WP) and electrons to pass the two discriminators are shown separately for \tauh and electrons with $20 < \pt < 100\GeV$ and $\pt > 100\GeV$.
Within the range of applicability of the previously used MVA discriminator, the \De discriminator increases the \tauh efficiency by consistently more than 10\% for a constant misidentification probability.
For example, the \tauh efficiency increases from 87 to 99\% for the loosest working point of the two discriminators with an electron efficiency of around 7\% for $\pt < 100\GeV$.
Besides increasing the reach in \tauh identification efficiency, the new discriminator also provides the possibility to reject electrons with an efficiency of $10^{-4}$ for a \tauh efficiency of 55\%.
Similar, albeit slightly smaller, gains are observed for $\pt > 100\GeV$.

\begin{figure}[ht]
	\centering
	\includegraphics[width=0.48\textwidth]{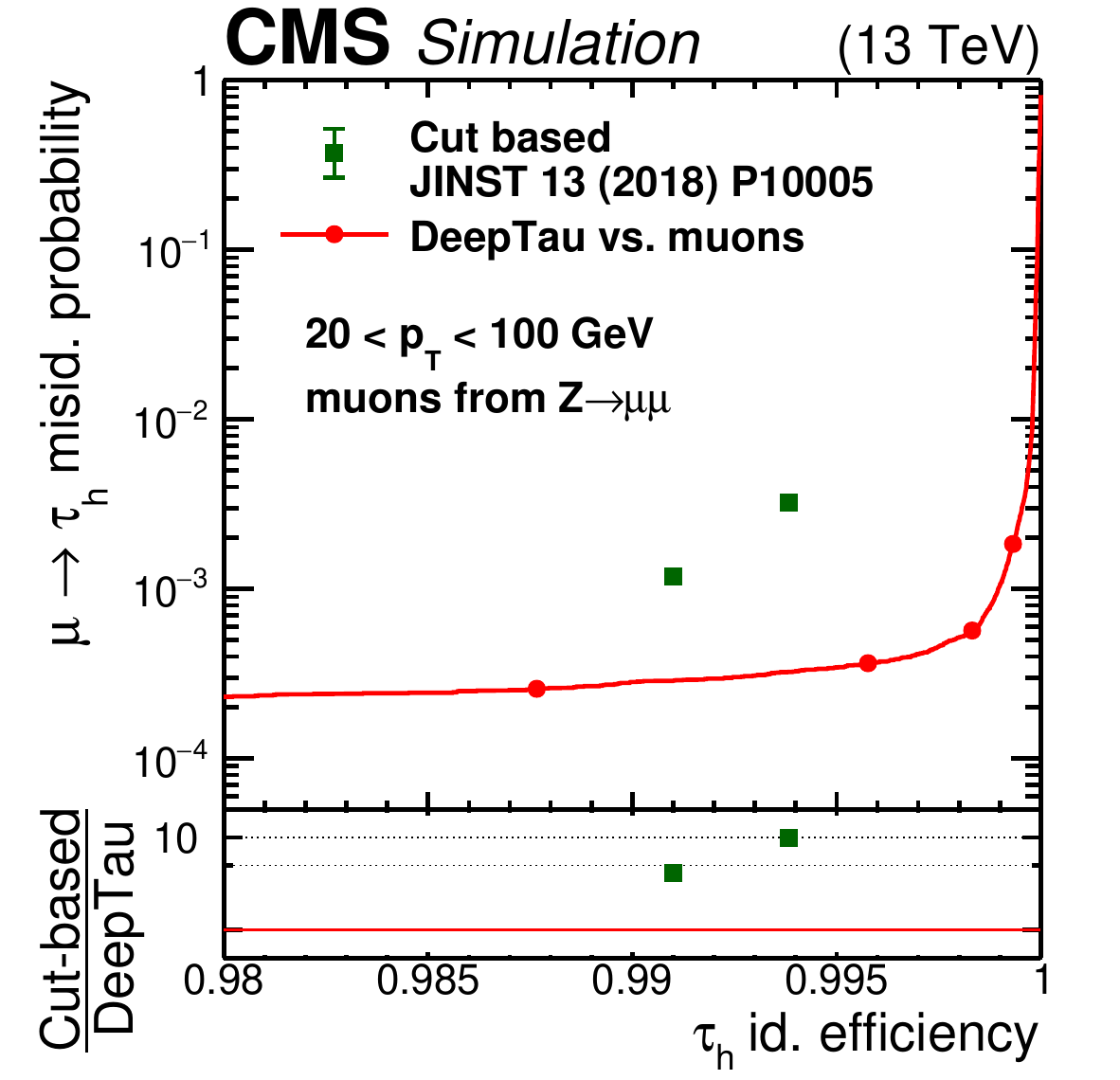}
	\includegraphics[width=0.48\textwidth]{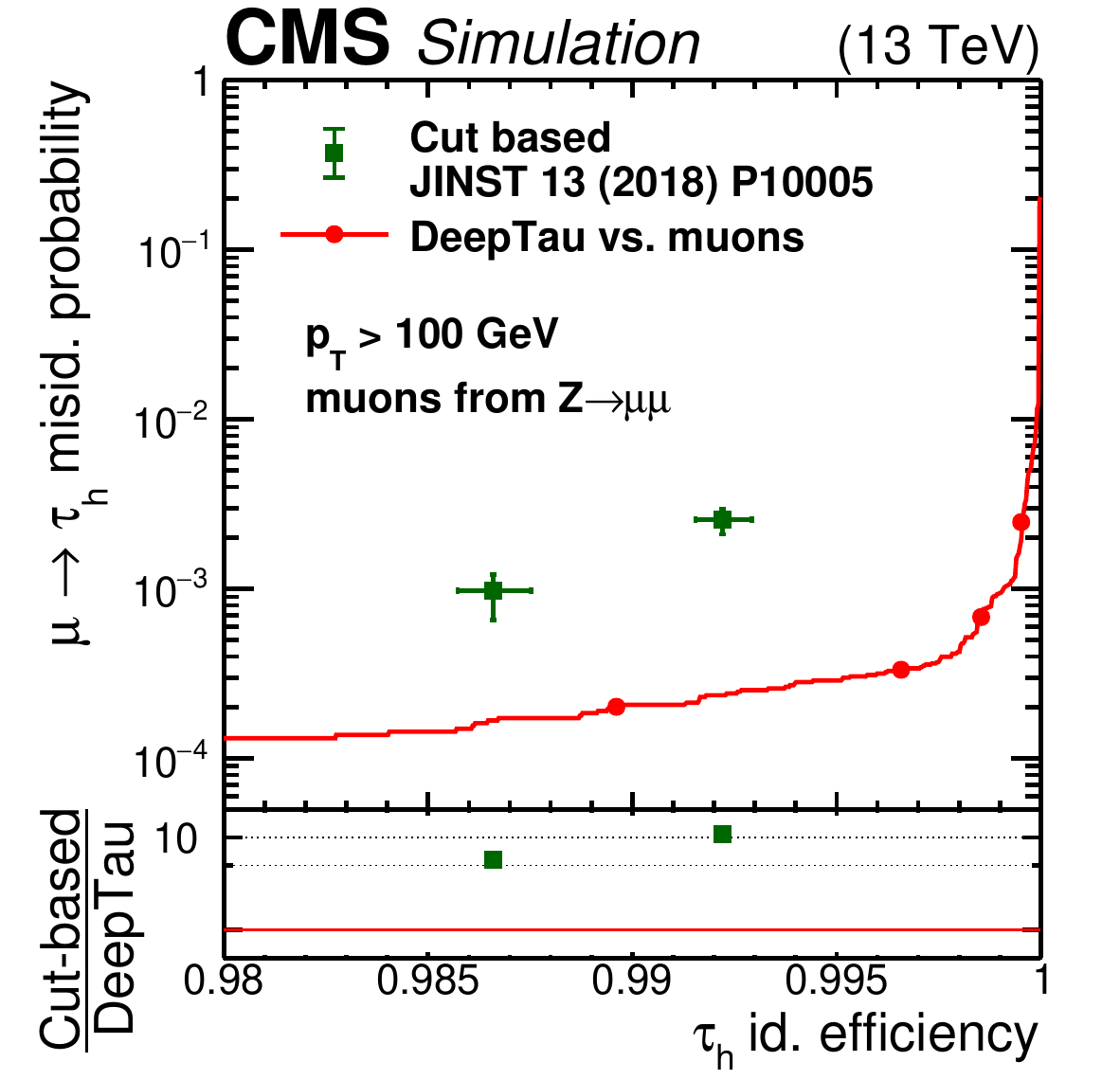}
	\caption{Efficiency for muons versus efficiency for simulated \tauh to pass the cutoff-based and \Dm discriminators, separately for muons and \tauh with $20 < \pt < 100\GeV$ (left) and $\pt > 100\GeV$ (right). The four working points are indicated as full circles. Vertical bars correspond to the statistical uncertainties. In both \pt regimes, the \Dm discriminator rejects up to a factor of 10 more muons at \tauh efficiencies around 99\%, and it leads to an increase of the \tauh efficiency for a similar background rejection by about 0.5\%.}
	\label{fig:rocs_muons}
\end{figure}

Gains are also observed in terms of the discrimination of \tauh candidates (with $\pt > 20\GeV$, $\abs{\eta} < 2.3$ and passing the VVVLoose \Djet and VVVLoose \De WP) versus muons, as shown in Fig.~\ref{fig:rocs_muons}.
Compared with the cutoff-based discriminator, the \Dm discriminator leads to an increase of the \tauh efficiency of around 0.5\% for a given prompt muon efficiency.
For the \tauh efficiency range of 99.1--99.4\% addressed by the cutoff-based discriminator, the \Dm discriminator leads to a factor of 3--10 larger prompt muon rejection, and thereby provides a significantly improved rejection of prompt muons for analyses with hadronically decaying tau leptons, while leading to only a percent-level loss of \tauh identification efficiency.

The performance results discussed so far have been obtained with simulated samples that are consistent with the 2017 data-taking conditions.
Moreover, after applying the discriminators to simulated samples that correspond to the 2016 and 2018 data-taking conditions, we see consistent performance, both in absolute terms and relative to other discriminators.
Based on these findings, the discriminators \De, \Dm, and \Djet have been recommended for all CMS analyses of LHC data at $\sqrt{s} = 13\TeV$.

To summarize, the \textsc{DeepTau} discriminators against jets, electrons, and muons lead to large gains in discrimination performance against various backgrounds.
The gains in \tauh identification efficiency at a typical jet or electron efficiency used in physics analyses amount to more than 30\% and 10\%, respectively.
Therefore, a large gain in reach is expected in physics analyses with tau leptons, in particular for final states with two \tauh candidates.
These improvements are confirmed by physics analyses that already used the \textsc{DeepTau} discriminators~\cite{CMS:2020dvp,Sirunyan:2020icl,Sirunyan:2020zbk}.

\section{Performance with \texorpdfstring{$\sqrt{s}=13\TeV$}{13 TeV} data}
\label{sec:performance}

The performance of the \tauh reconstruction and of the \textsc{DeepTau} discriminators is validated with collision data at $\sqrt{s} = 13\TeV$.
The total integrated luminosity for the 2016--2018 data collected with the CMS detector and validated for use in analysis amounts to 138\fbinv, of which 36\fbinv were recorded in 2016, 42\fbinv in 2017, and 60\fbinv in 2018~\cite{CMS:2021xjt,CMS-PAS-LUM-17-004,CMS-PAS-LUM-18-002}.
The reconstruction and identification efficiencies for \tauh are measured separately for the three data-taking years using a $\PZ\to\Pgt\Pgt$ event sample in the $\Pgm\tauh$ final state, following methods similar to those established previously~\cite{Sirunyan:2018pgf}.
In addition, measurements are made of the misidentification efficiency with which jets, electrons, and muons are reconstructed and identified as \tauh candidates.
All of these measurements are essential ingredients for physics analyses that use \tauh candidates.

In this paper, we present a subset of the measurements that were carried out, mostly based on the largest data set recorded in 2018.
The identification efficiencies are generally measured for all working points introduced above.
We discuss a subset of measurements here only for representative working points. 
Consistent results have been obtained for the other working points, and for the 2016 and 2017 data sets.
We have not performed such efficiency measurements for the decay modes with missing charged hadrons; future analyses including these decay modes will need to derive appropriate corrections.
The full set of efficiency and energy scale corrections together with their corresponding uncertainties is available for data analyses in the CMS Collaboration.

\subsection{Reconstruction and identification efficiency}

The measurement of the \tauh reconstruction and identification efficiencies uses a $\Pgm\tauh$ event sample that is selected in the same way as in the previous measurement~\cite{Sirunyan:2018pgf}, with the following updates.
Events are recorded with a single-muon trigger with a nominal \pt threshold of 24\GeV and are required to have at least one reconstructed and isolated muon with $\pt > 25\GeV$.
The \tauh candidate is required to be reconstructed with the decay mode finding algorithm, to have $\pt > 20\GeV$, and to pass a given threshold on the \Djet discriminator.
In addition, the \tauh candidate needs to pass the VVLoose working point of the \De discriminator and the Tight working point of the \Dm discriminator, to reject background from muons or electrons misidentified as a \tauh candidate.
Additional criteria are applied to increase the purity of events with genuine \tauh candidates, including a requirement on the transverse mass of the muon and \ptvecmiss ($\sqrt{2\pt^\mu\ptmiss(1 - \cos\Phi)} < 60\GeV$, where $\Phi$ is the angle in the transverse plane between the muon momentum and \ptvecmiss) and a maximum difference in $\eta$ between the reconstructed muon and the \tauh candidate ($\abs{\Delta\eta} < 1.5$).
The resulting dataset is enriched in $\PZ\to\Pgt\Pgt$ events, with the residual contamination from other processes amounting to approximately one fifth of the total yield (Fig.~\ref{fig:data_validation}).
In addition to the $\Pgm\tauh$ event sample, a $\Pgm\Pgm$ event sample is defined to normalize the $\PZ\to\Pgt\Pgt$ event yields.
This event sample is subject to the same trigger and muon selection criteria as the $\Pgm\tauh$ event sample, such that related uncertainties partially cancel in the normalization scale factor.

The efficiencies are extracted from a maximum likelihood fit to the distribution of the reconstructed visible invariant mass of the $\Pgm\tauh$ system, \mvis, and to the expected and observed event yields in the $\Pgm\Pgm$ region.
In the fit, the ratio of the observed and expected efficiency for a genuine \tauh to pass the selection is incorporated as a free parameter, such that the fit yields a scale factor with respect to the simulated efficiency.
All known sources of systematic uncertainties are incorporated into the fit. 
Uncertainties include the uncertainty in the integrated luminosity, uncertainties in the muon trigger, identification, and isolation efficiencies,  uncertainties due to the limited number of simulated events, and uncertainties in the normalizations of \ttbar, QCD multijet, and $\PZ/\gamma^*$+jets background.
Since the scale factors are extracted after the full event selection, they incorporate any differences between data and simulation of the full product of all reconstruction and identification efficiencies.
The scale factors are extracted from the maximum likelihood fit together with their corresponding uncertainties using the asymptotic properties of the likelihood function~\cite{Cowan:2010js}.

\begin{figure}[ht]
	\centering
	\includegraphics[width=0.48\textwidth]{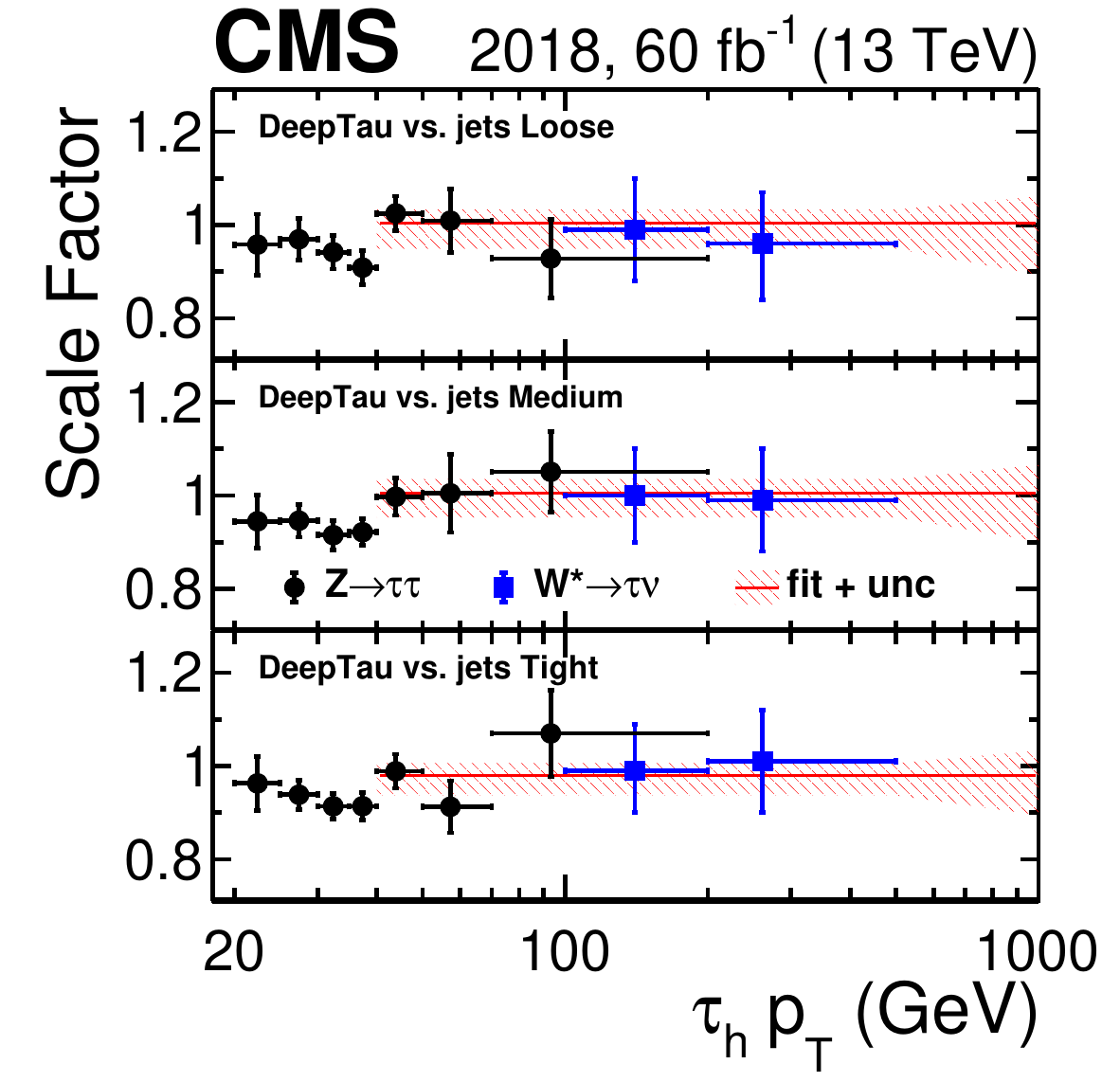}
	\includegraphics[width=0.48\textwidth]{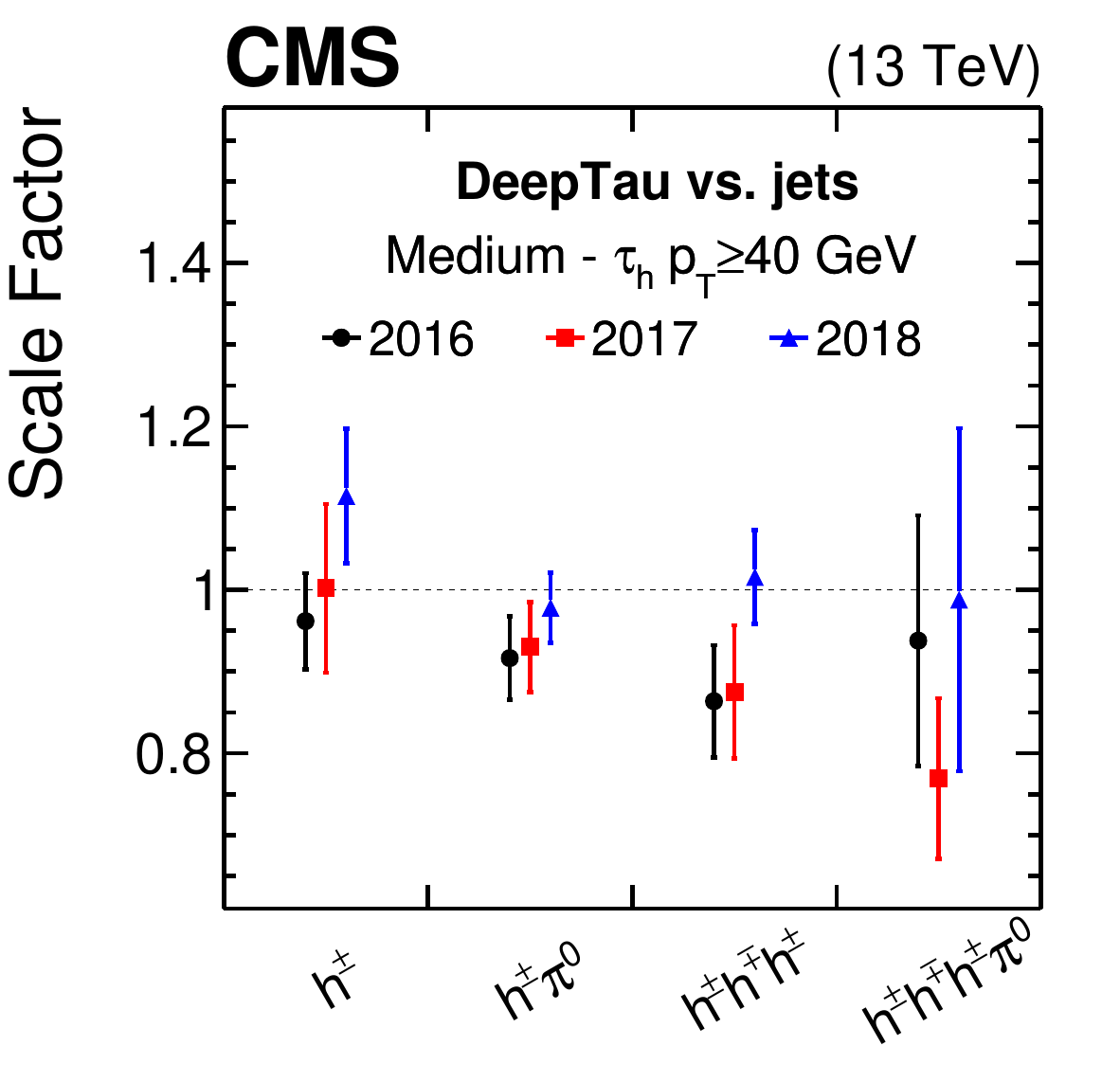}
	\caption{Data-to-simulation scale factors for genuine \tauh to be reconstructed as \tauh candidates and to pass the Loose, Medium, and Tight working points of the \Djet discriminator as a function of the \tauh candidate \pt (left). Vertical bars correspond to the combined statistical and systematic uncertainties in the individual scale factors. The red hatched bands indicate the uncertainties for $\pt > 40\GeV$, obtained from a combination of the individual measurements.
	The right plot shows the data-to-simulation scale factors for the \tauh candidates with $\pt > 40\GeV$ to pass the Medium \Djet working point as a function of reconstructed \tauh decay mode. The efficiencies are obtained with 2018 data and the according simulated events using a likelihood fit to the distribution of the reconstructed $\mvis(\Pgm, \tauh)$. The scale factors are shown separately for data taken in 2016, 2017, and 2018 (and the corresponding simulated events) and for the four main \tauh decay modes.}
	\label{fig:sf_tau}
\end{figure}

The scale factors are extracted in different \pt bins, $\pt\in\{ [20, 25]$, $[25, 30]$, $[30, 35]$, $[35, 40]$, $[40, 50]$, $[50, 70]$, ${>}70\GeV\}$, to account for a possible \pt dependence of the extracted scale factors, which would affect analyses that use different minimum \pt thresholds or that analyze \tauh final states from processes with different \pt distributions.
While analyses usually use all reconstructed decay modes, except for the modes with missing charged hadrons so a parametrization in terms of decay modes is not necessary, analyses that select events using the di-\tauh trigger algorithms observe a different proportion of \tauh candidates reconstructed in the various decay modes.
This is a consequence of the dependency of \tauh trigger efficiency on the various decay modes.
Therefore, scale factors are separately determined in four bins of reconstructed decay modes (\oneProngZeroPizero; \oneProngOnePizero or \oneProngTwoPizero; \threeProngZeroPizero; and \threeProngOnePizero).
These scale factors are shown in Fig.~\ref{fig:sf_tau} for \tauh $\pt > 40\GeV$, which corresponds to the minimum recommended threshold in analyses that use events recorded with the di-\tauh trigger.

Since the reach in \tauh \pt is limited in the $\PZ\to\Pgt\Pgt$ sample, a separate off-shell \PW boson ($\PW^*$) event sample is used to measure additional scale factors for \tauh $\pt > 100\GeV$ in a \tauh-plus-\ptmiss final state without an additional hadronic jet, following closely the method discussed in Ref.~\cite{Sirunyan:2018pgf}.
The requirement of high \tauh \pt and high \ptmiss combined with the hadronic jet veto leads to an event sample dominated by off-shell $\PW^*\to\Pgt\Pgngt$ boson decays.
The events are recorded with a trigger that requires a large \ptmiss.
The efficiencies are obtained from a maximum likelihood fit to the transverse mass distribution of the \tauh candidate and \ptmiss.
To constrain the fiducial $\PW^*$ cross section in the signal region, a $\PW^*\to\Pgm\Pgngm$ control region is also included in the fit.

Figure~\ref{fig:sf_tau} (left) shows the scale factors as a function of the reconstructed \tauh candidate \pt separately for the Loose, Medium, and Tight working points.
The plots also show the uncertainties for the \tauh candidates with $\pt > 40\GeV$, which are obtained from a constant fit up to \pt values of 500\GeV, assuming there is no statistically significant \pt dependence of the scale factors.
Beyond 500\GeV, the uncertainties are enlarged because of the lack of \tauh candidates with such high \pt values, with the uncertainty at 1000\GeV twice as large as at 500\GeV.

The \pt-dependent uncertainties in the \tauh scale factors range from 2 to 5\%.
For the \tauh \pt spectrum in the $\PZ\to\Pgt\Pgt$ sample, the combined scale factor uncertainty amounts to ${\approx}$2\%.
These uncertainties are considerably smaller than in previous measurements, which reported combined uncertainties of ${\approx}$6\%~\cite{Sirunyan:2018pgf}.
The previous measurements used the ``tag-and-probe'' technique~\cite{CMS:2010svw} that only measured the identification efficiency for the discriminator under consideration and added additional uncertainties in the reconstruction efficiency from independent measurements of single charged-hadron reconstruction efficiencies.
The measurement of the product of all reconstruction and identification efficiencies with a maximum likelihood fit hence represents another important improvement for physics analyses with \tauh final states, as it leads to a considerable reduction of the systematic uncertainties related to the \tauh reconstruction and identification.

The data-to-simulation scale factors as a function of the \tauh decay mode are shown in Fig.~\ref{fig:sf_tau} (right) for \tauh $\pt > 40\GeV$ for the three data-taking years (2016, 2017, and 2018).
The scale factors are generally a bit smaller than one but consistent with unity within 10\% (20\% for the \tauh candidates reconstructed in the \threeProngOnePizero decay mode).
The scale factors below one can be traced back to different origins, including the imperfect modelling of hadronization and an imperfect simulation of the detector alignment and track hit reconstruction efficiencies.

\subsection{The \texorpdfstring{\tauh}{hadronic tau decay} energy scale}

The \tauh energy scale is obtained from the same $\Pgm\tauh$ event samples as the scale factors discussed in the previous section, following the same method described in previous publications~\cite{Sirunyan:2018pgf}.
For the main measurement discussed here, \tauh candidates are required to pass the Medium working point of the \Djet discriminator.
Consistent results are obtained with cross-check measurements for various working points of the \Djet discriminator.
To improve the a priori data-simulation agreement, scale factors as described in the previous section are already applied to simulated samples with \tauh final states.

Two different observables are used to measure the \tauh energy scale:
the $\Pgm\tauh$ invariant mass (\mvis) and the reconstructed \tauh mass ($m(\tauh)$).
For \tauh candidates reconstructed in the \oneProngZeroPizero decay mode, $m(\tauh)$ is constant, so the measurement is only performed with the \mvis observable.

Simulated templates of the \mvis and $m(\tauh)$ distributions are created for \tauh energy scale shifts between $\pm 3\%$ in steps of 0.2\%.
For each shift, a maximum likelihood fit is performed, using the combined expectation from the $\PZ\to\Pgt\Pgt$ in the $\Pgm\tauh$ final state and the background events, using the set of systematic uncertainties discussed above.
The observed value of the \tauh energy scale shift is obtained from the distribution of the likelihood ratio together with the corresponding uncertainty.

\begin{figure}[h!t]
	\centering
	\includegraphics[width=0.65\textwidth]{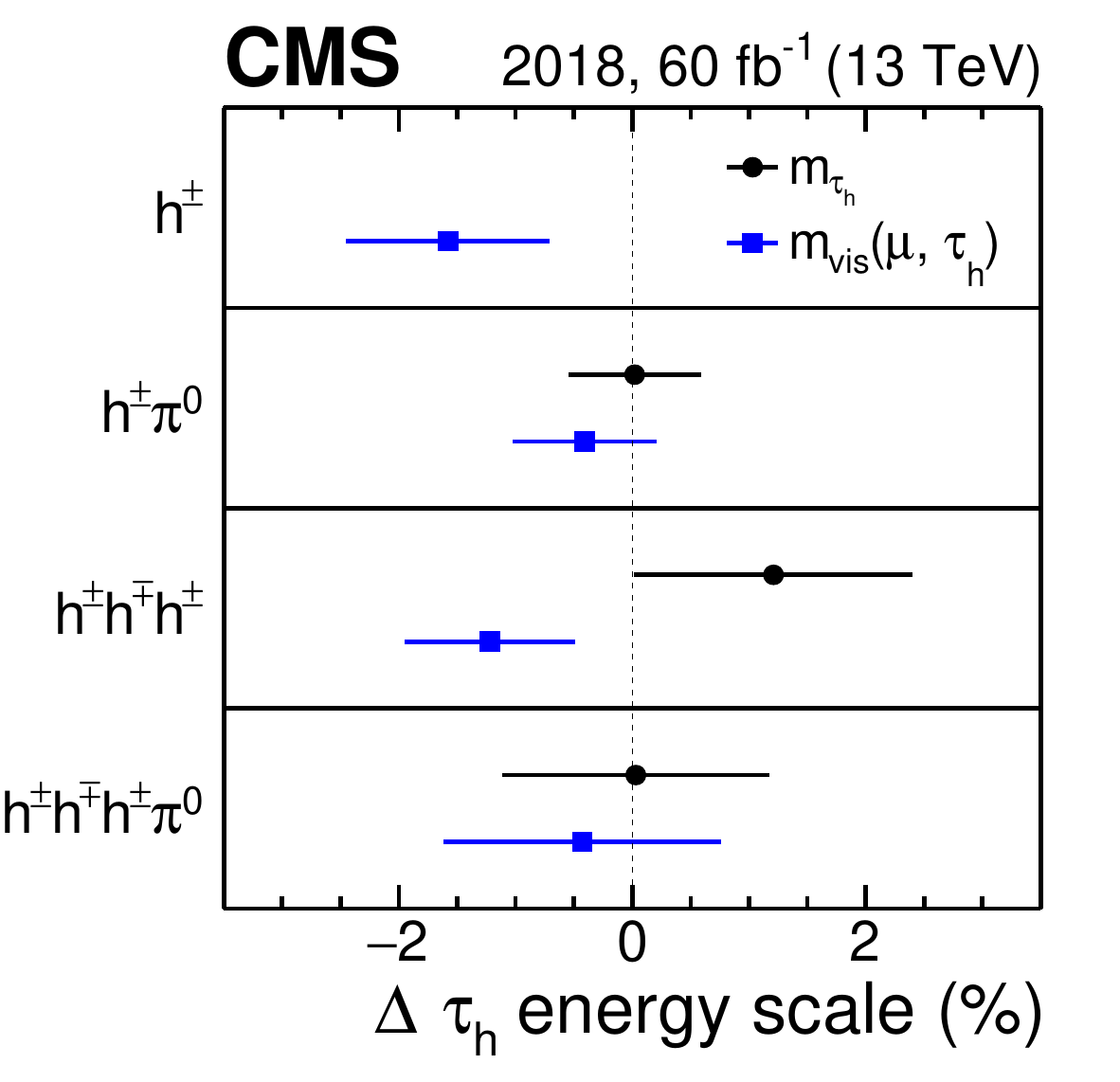}
	\caption{Relative difference between \tauh energy obtained in data and simulated events for the four main reconstructed \tauh decay modes for the 2018 data set. The results are obtained from fits to the distribution of either the reconstructed $\mvis(\Pgm, \tauh)$ (blue lines) or \mtauh (black lines). The horizontal bars represent the uncertainties in the measurements. The measured values are consistent with no shift of the \tauh energy scale between data and simulation, with the largest difference amounting to 1.5 standard deviations.}
	\label{fig:energy_scale}
\end{figure}

The resulting relative differences between the \tauh energy scale in data and simulation are shown in Fig.~\ref{fig:energy_scale} separately for the two different measurements.
The relative difference is given as a correction, in percent, to a unit multiplicative factor applied to the \tauh four-momentum in the simulation.
The results obtained with the two methods are consistent with each other, and the \tauh energy scale shifts are consistent with no data-simulation differences, with the largest difference amounting to 1.5 standard deviations.
The relative uncertainties range from 0.6\% for the \oneProngOnePizero and \threeProngZeroPizero decay modes to 0.8\% for the \oneProngZeroPizero mode.

\begin{figure}[h!t]
	\centering
	\includegraphics[width=0.48\textwidth]{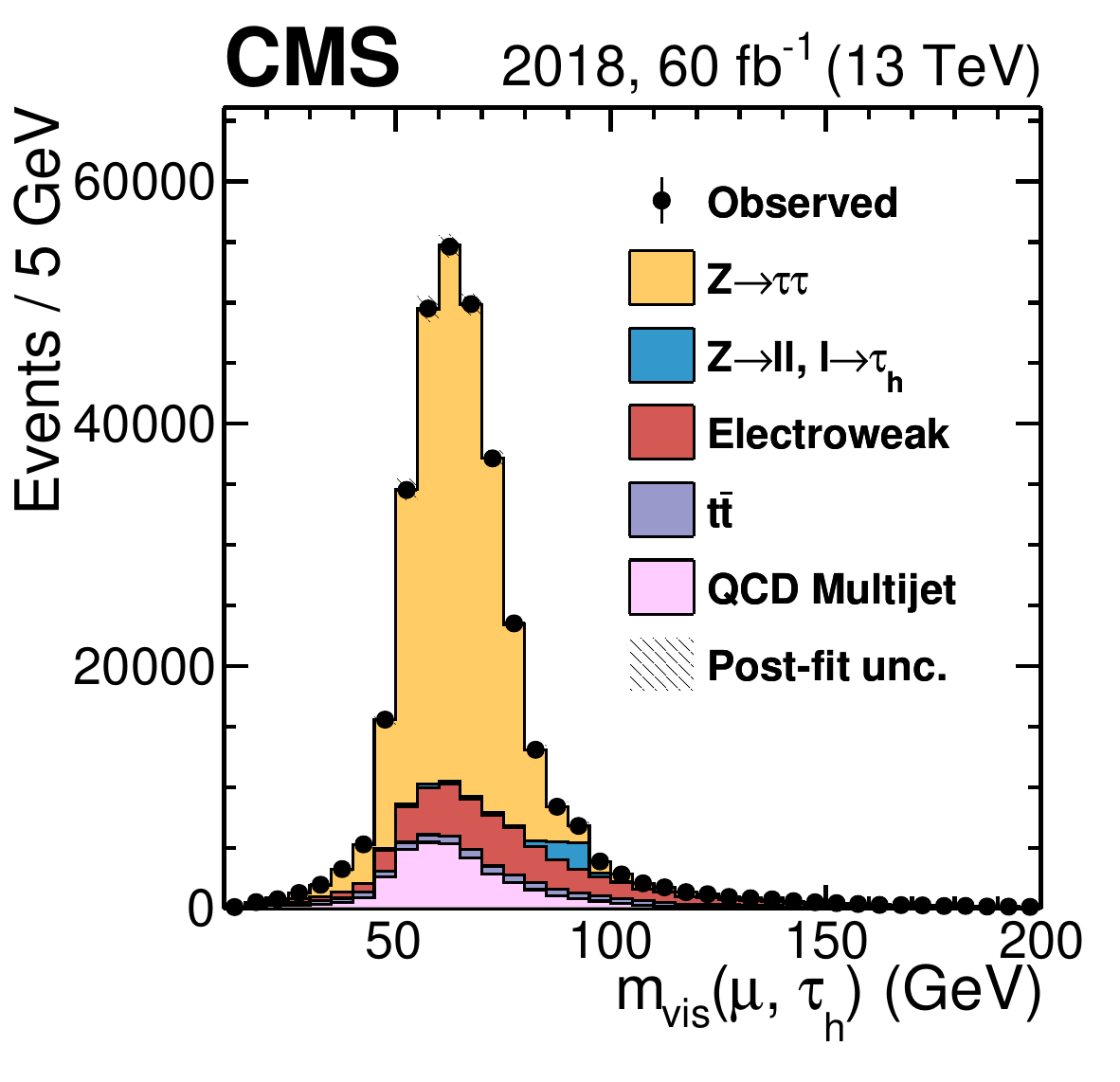}
	\includegraphics[width=0.48\textwidth]{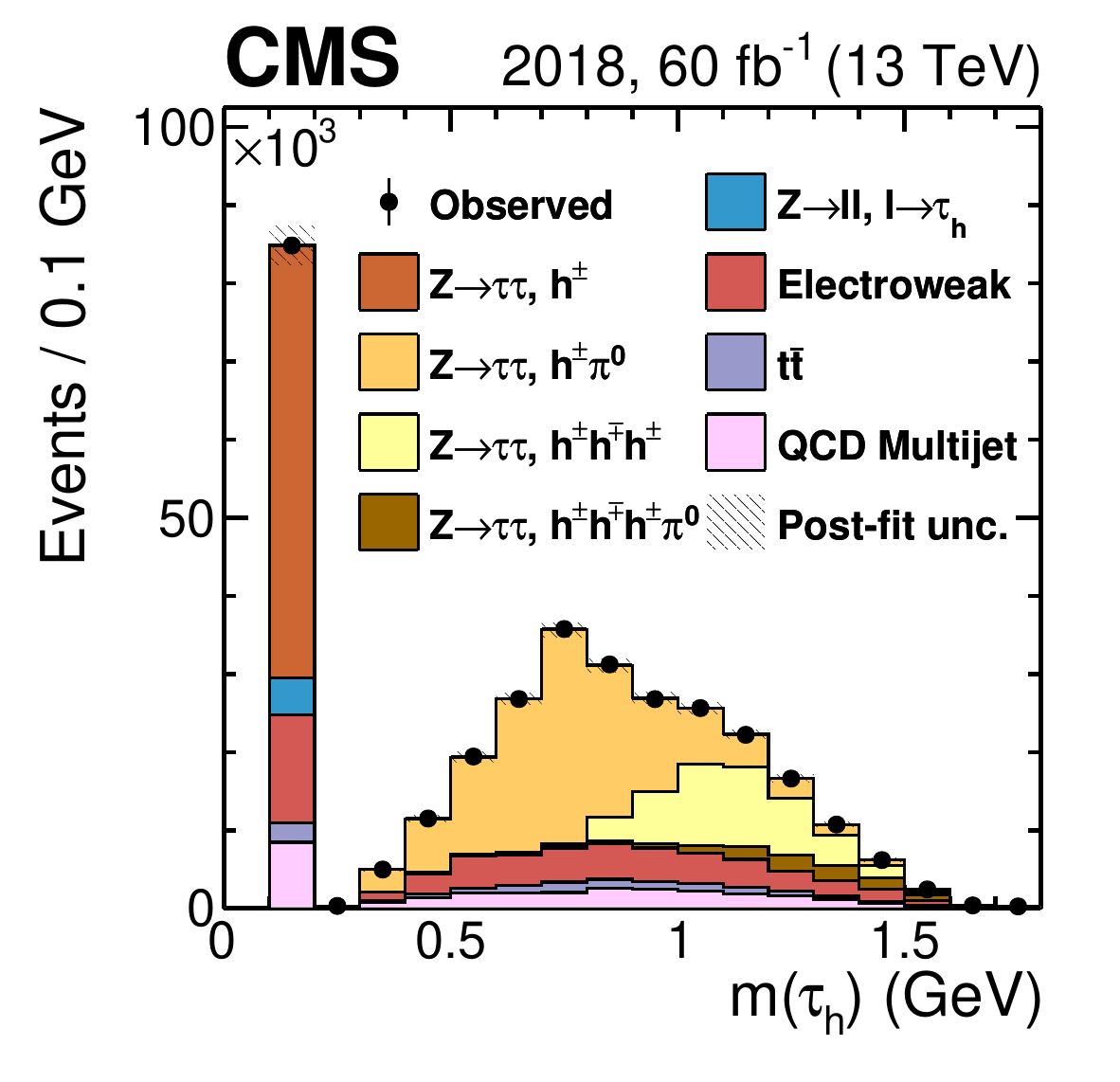}
	\caption{Distribution of the reconstructed visible invariant mass of the $\Pgm\tauh$ system, $\mvis(\Pgm, \tauh)$ (left) and of the visible invariant \tauh mass (right). Vertical bars correspond to the statistical uncertainties. The event selection corresponds to the one in the measurement of the \tauh reconstruction and identification efficiencies and uses the Tight working point of the \Djet discriminator. The distributions incorporate all measured scale factors and energy corrections and are scaled to the outcome of a maximum likelihood to the observed data with the $\PZ\to\Pgt\Pgt$ contribution freely floating. The electroweak background combines contributions from single top quark, diboson, and $\PW$+jets processes as well as $\PZ(\to\Pgt\Pgt)$+jets events where the reconstructed \tauh originates from a jet misidentified as a \tauh candidate instead of one of the produced tau leptons. In the $m(\tauh)$ distribution, the $\PZ\to\Pgt\Pgt$ contributions are shown separately for the different \tauh decay modes.}
	\label{fig:data_validation}
\end{figure}

To validate the extracted scale factors and to illustrate the distributions used for the extraction of the scale factors and the \tauh energy corrections, distributions of \mvis and $m(\tauh)$ are investigated, requiring the \tauh to pass the Tight working point of the \Djet discriminator (Fig.~\ref{fig:data_validation}).
These distributions are obtained from a maximum likelihood fit to the $\Pgm\Pgt$ data, applying the same selection used in the extraction of the scale factors.
This fit incorporates the derived scale factors and energy corrections with their uncertainties, together with the full set of systematic uncertainties used in the scale factor extractions.
In the fit to data, the $\PZ\to\Pgt\Pgt$ normalization is a freely floating parameter.
Data and predictions agree within the uncertainties, indicating that the derived scale factors and energy corrections lead to a good and complete description of the data, including the description of the different \tauh decay modes.

\subsection{Lepton and jet misidentification efficiencies}

The efficiencies for electrons, muons, and jets to pass the respective \textsc{DeepTau} discriminators are measured with dedicated event samples, following closely the measurements explained in detail in Ref.~\cite{Sirunyan:2018pgf}.
The efficiencies for electrons and muons to be reconstructed and misidentified as \tauh candidates are measured with $\PZ\to\Pell\Pell$ events ($\Pell\in\{\Pe, \Pgm\}$) where one \Pell is misreconstructed as a \tauh candidate; the event is reconstructed as an $\Pell\tauh$ final state.
The efficiencies are extracted with the tag-and-probe method, where \Pell candidates that pass a given working point of the \Dm or \De discriminator end up in the pass region and those that do not pass in the fail region.
A maximum likelihood fit is performed to obtain the efficiencies. 
The distribution used in the pass region is the invariant $\Pell\tauh$ mass (\mvis), whereas the fail region is included as a single bin.
Similar to the extraction of the \tauh efficiencies, all relevant uncertainties are included in the maximum likelihood fit.
To reject background from jets reconstructed as \tauh candidates, the candidates are also required to pass the Medium working point of the \Djet discriminator.
An additional parameter is introduced in the fit that scales the contributions in the pass and fail regions simultaneously and corresponds to the scale factor for the efficiency to pass the \Djet discriminator.
This approach is based on the assumption that the probability for electrons or muons to pass the \Djet discriminator is independent of the probability to pass the \De or \Dm discriminator.
Another parameter is introduced for electrons misreconstructed as \tauh candidates that allows for a shifted energy scale.

\begin{figure}[h!t]
	\centering
	\includegraphics[width=0.48\textwidth]{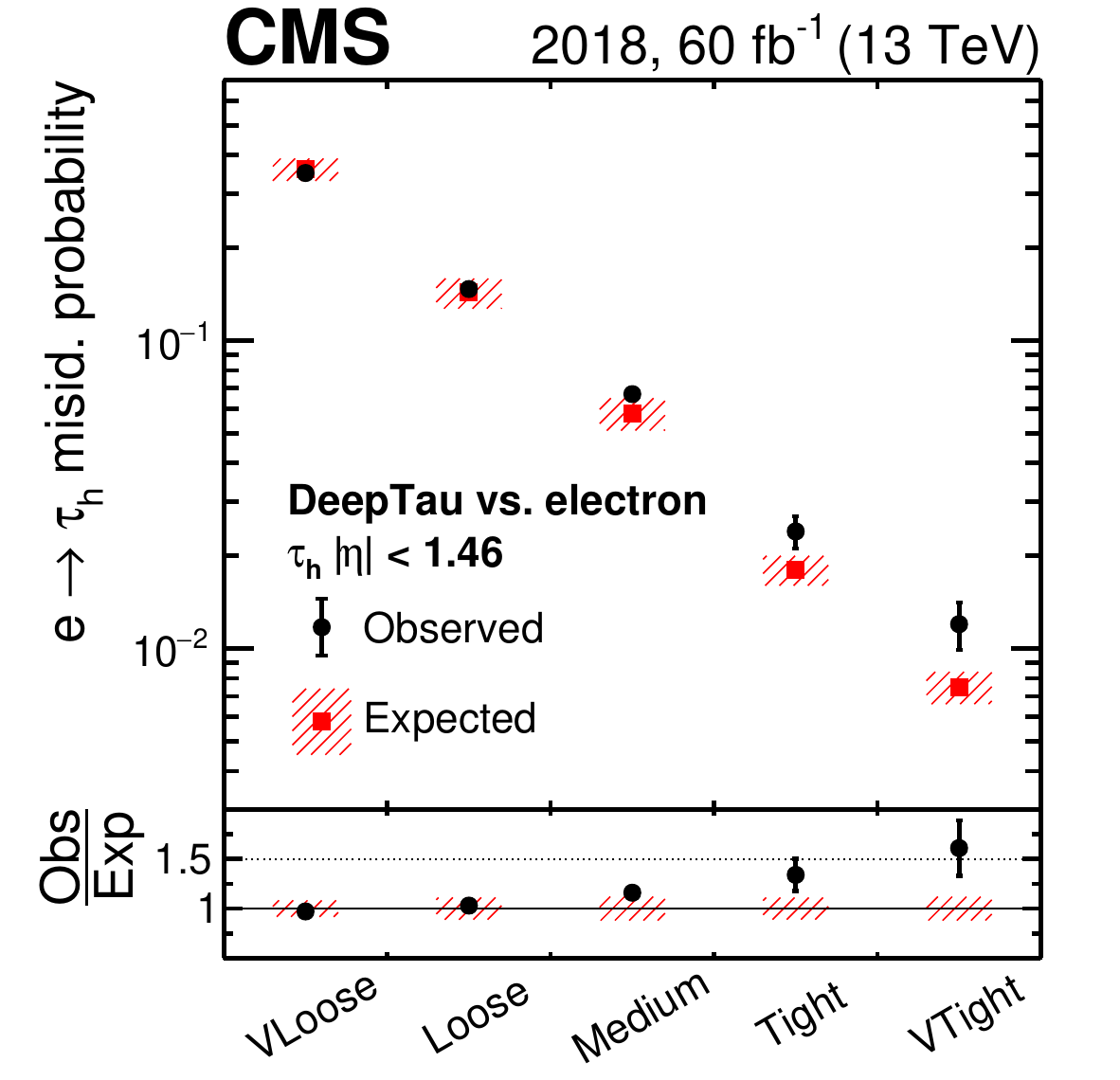}
	\includegraphics[width=0.48\textwidth]{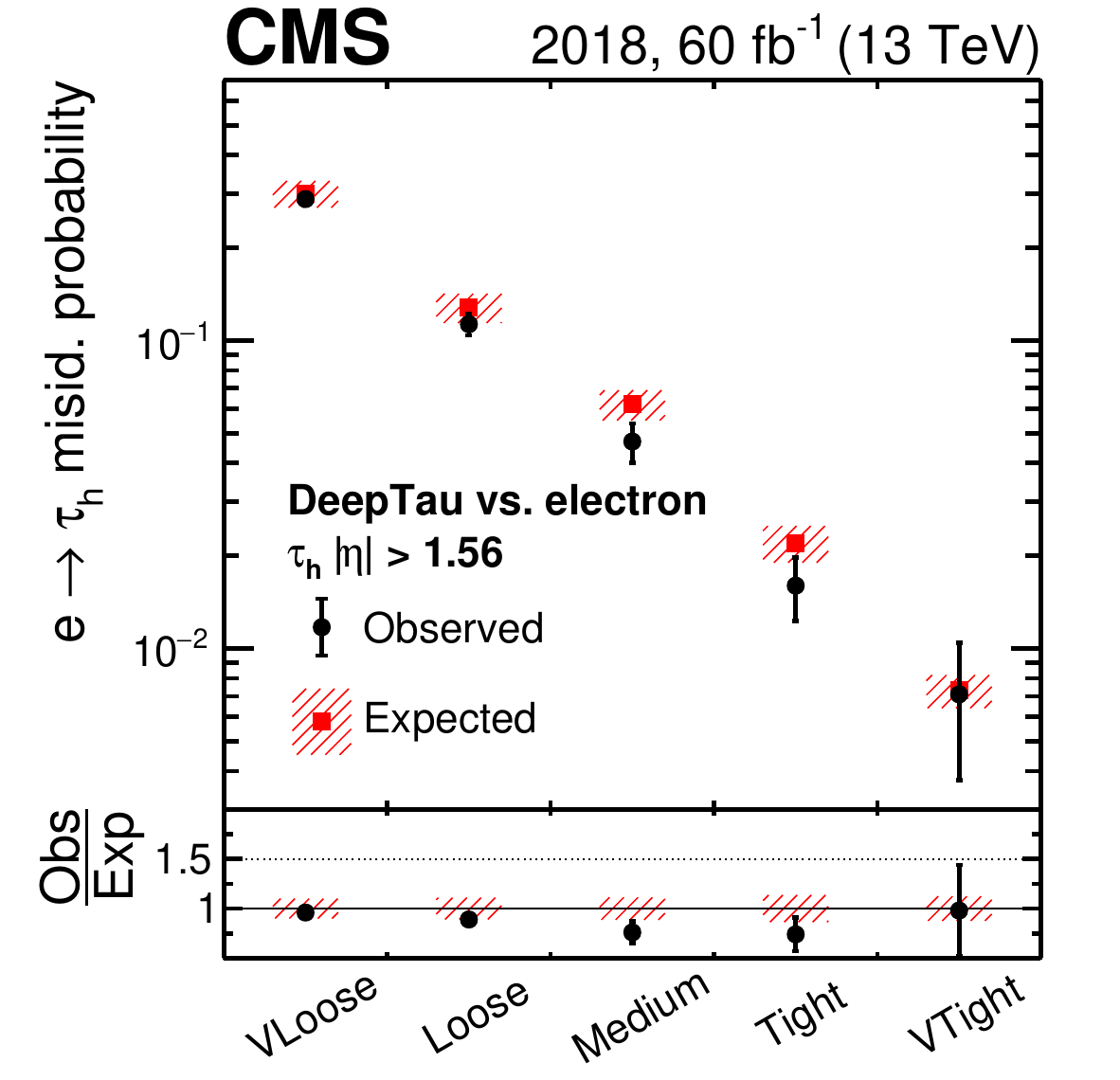}
	\caption{Observed and expected efficiencies for electrons to pass different \De working points. Vertical bars correspond to the combined statistical and systematic uncertainties in the individual scale factors. The electrons are required to be reconstructed as a \tauh candidate and to pass the medium \Djet working point. The efficiencies are shown separately for electrons with $\abs{\eta} < 1.46$ (left), corresponding to the ECAL barrel region, and $\abs{\eta} > 1.56$ (right), corresponding to the ECAL endcap regions.}
	\label{fig:sf_e}
\end{figure}

The $\Pe\tauh$ events for the measurement of the scale factors for electrons to pass different working points of the \De discriminator are recorded with a single-electron trigger with a \pt threshold of 35\GeV.
The efficiencies for electrons to pass the different working points in data and simulated events as well as the ratio of the efficiencies are shown in Fig.~\ref{fig:sf_e}.
The efficiencies are derived separately in the ECAL barrel region ($\abs{\eta} < 1.46$) and the ECAL endcap region ($\abs{\eta} > 1.56$).
The observed and expected efficiencies agree within a few percent for the Loose and VLoose working points in the barrel and within 11\% in the endcap region.
For tighter working points, scale factors are obtained that are significantly different from one and that are generally greater than one in the barrel, with a maximum scale factor of $1.61 \pm 0.28$ for the VTight working point, and smaller in the endcap.

\begin{figure}[h!t]
	\centering
	\includegraphics[width=0.48\textwidth]{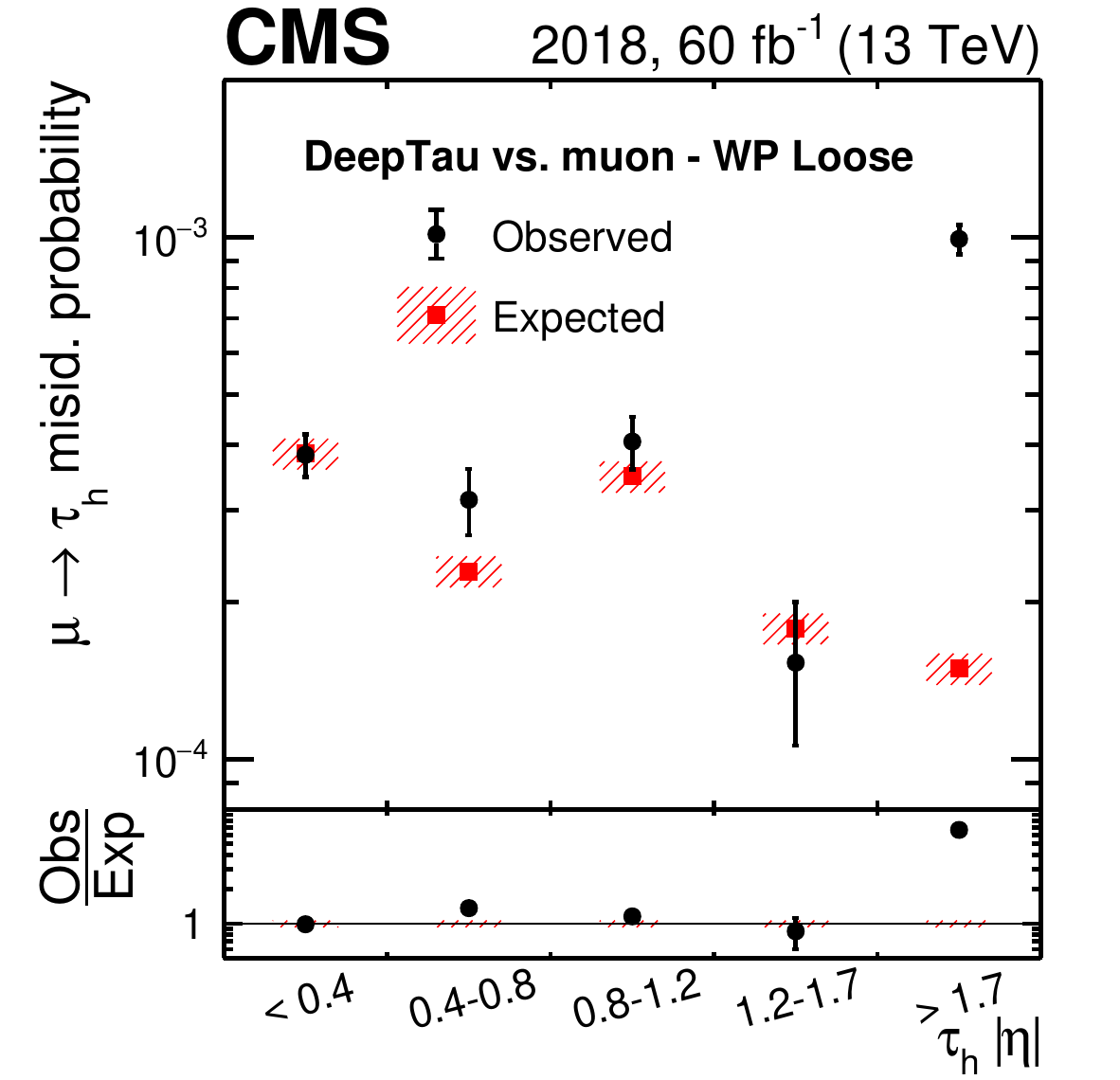}
	\includegraphics[width=0.48\textwidth]{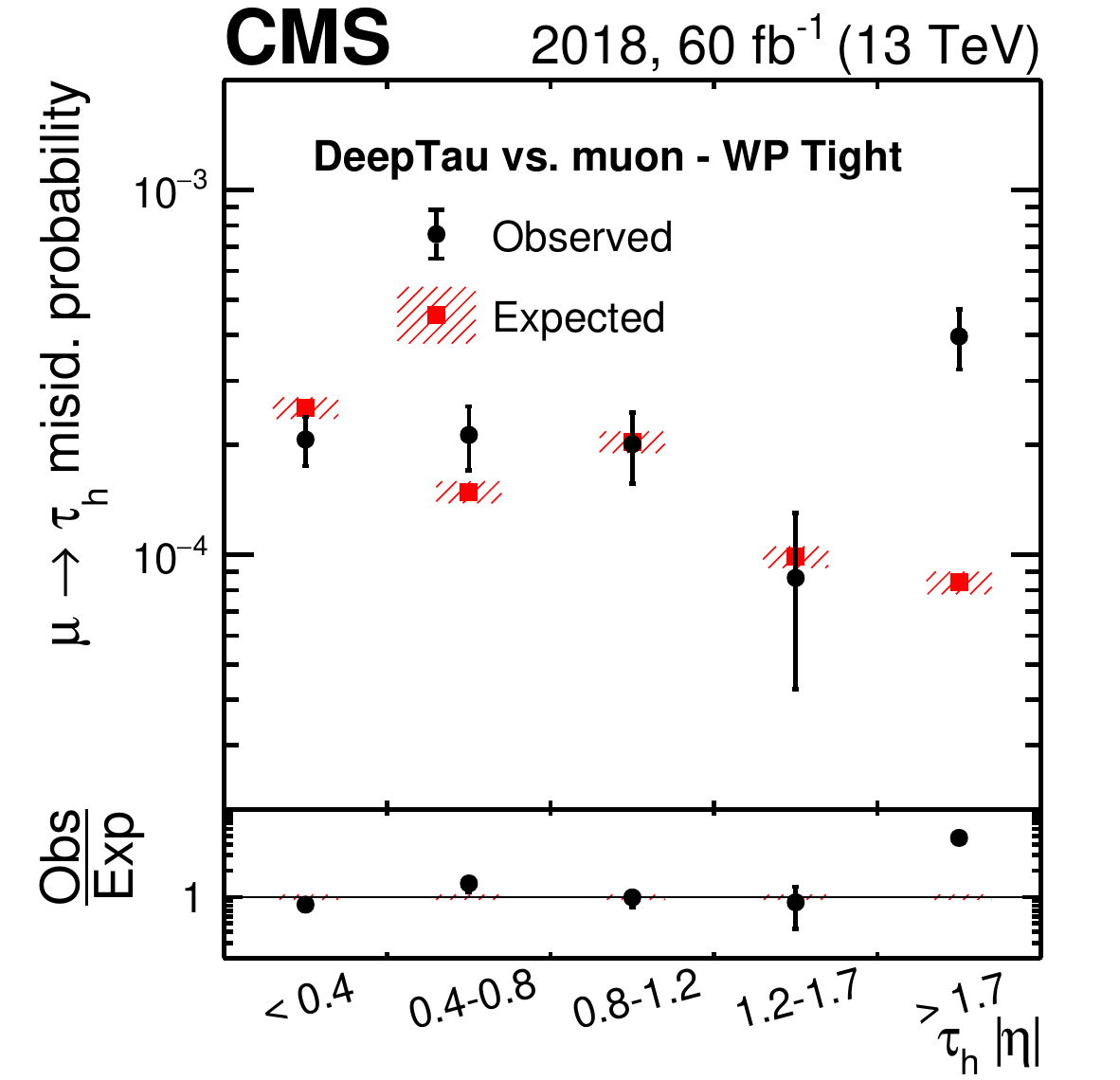}
	\caption{Observed and expected efficiencies for muons to pass the loose (left) and tight (right) \Dm working points. Vertical bars correspond to the combined statistical and systematic uncertainties in the individual scale factors. The muons are required to be reconstructed as \tauh candidates and to pass the Medium \Djet working point. The efficiencies are shown for several bins in $\abs{\eta}$.}
	\label{fig:sf_mu}
\end{figure}

The $\Pgm\tauh$ events for the measurement of the scale factors for muons to be misreconstructed and misidentified as \tauh candidates are recorded with a single-muon trigger with a threshold of 27\GeV.
The leading muon is required to have $\pt > 28\GeV$, and the transverse mass of the muon and \ptvecmiss is required to be smaller than 30\GeV.
The resulting data-to-simulation scale factors are extracted from a fit to the \mvis distribution and are shown in Fig.~\ref{fig:sf_mu}.
The scale factors are shown separately for the Loose and Tight working points, and for five different bins in $\abs{\eta}$.
Although the scale factors are consistent with each other for the different working points and consistent with unity for four of the $\abs{\eta}$ bins, they differ significantly from one for $\abs{\eta} > 1.7$. As a result of the high rejection power of \Dm, muons that are not discarded belong to atypical regions of the phase space, far from the bulk of the distributions.
For such uncommon muons, we observe that the simulation does not model the data well, especially for observables related to the muon track quality at large $\abs{\eta}$; this is compatible with being the cause of the sizeable data-to-simulation difference at $\abs{\eta} > 1.7$.
The large \Dm scale factors at $\abs{\eta} > 1.7$ have no notable impact on physics analyses because of the small probability for muons to pass any \Dm working point and the limited extent of the region presenting large data-to-simulation differences.

\begin{figure}[h!t]
	\centering
	\includegraphics[width=0.48\textwidth]{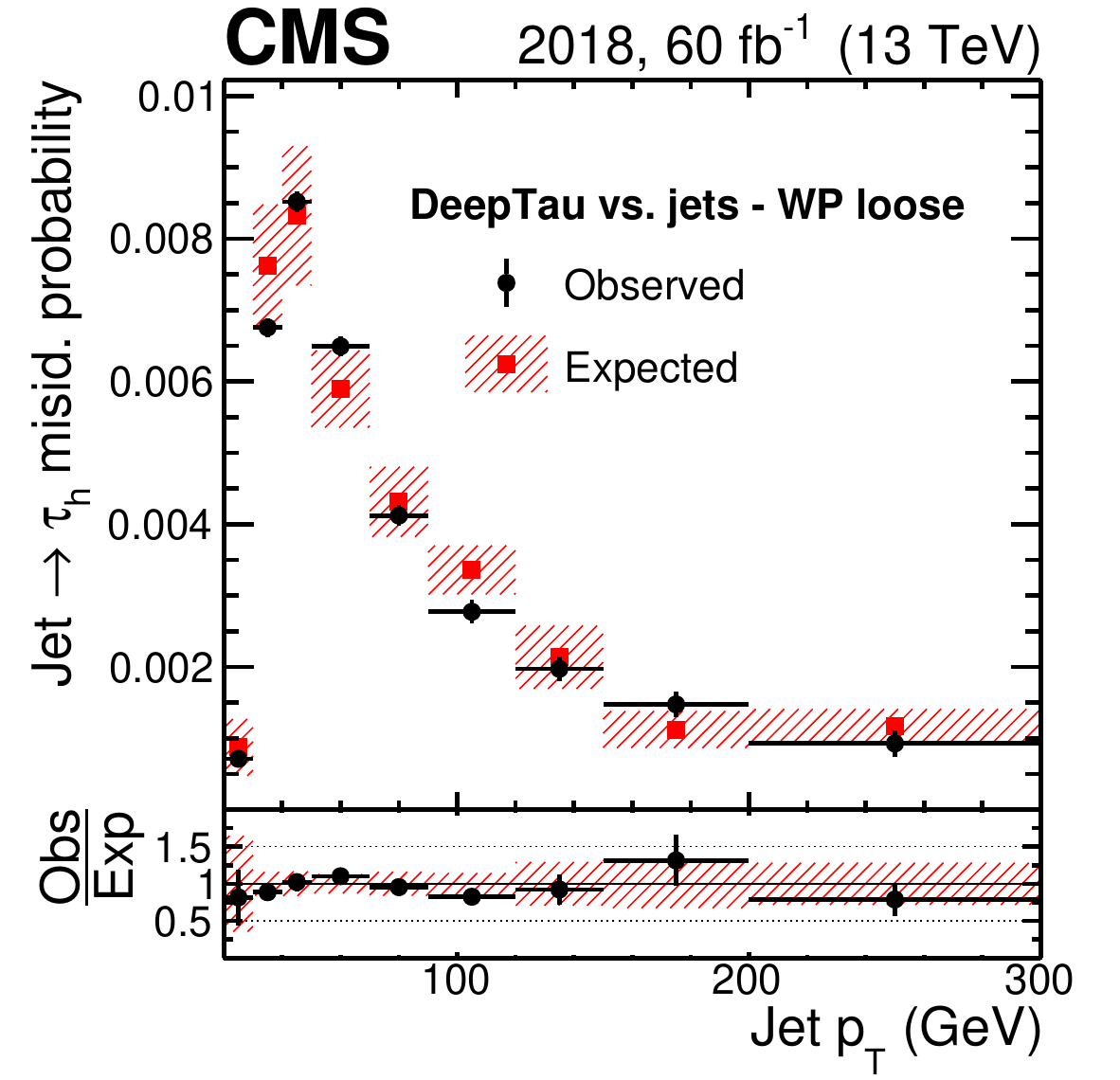}
	\includegraphics[width=0.48\textwidth]{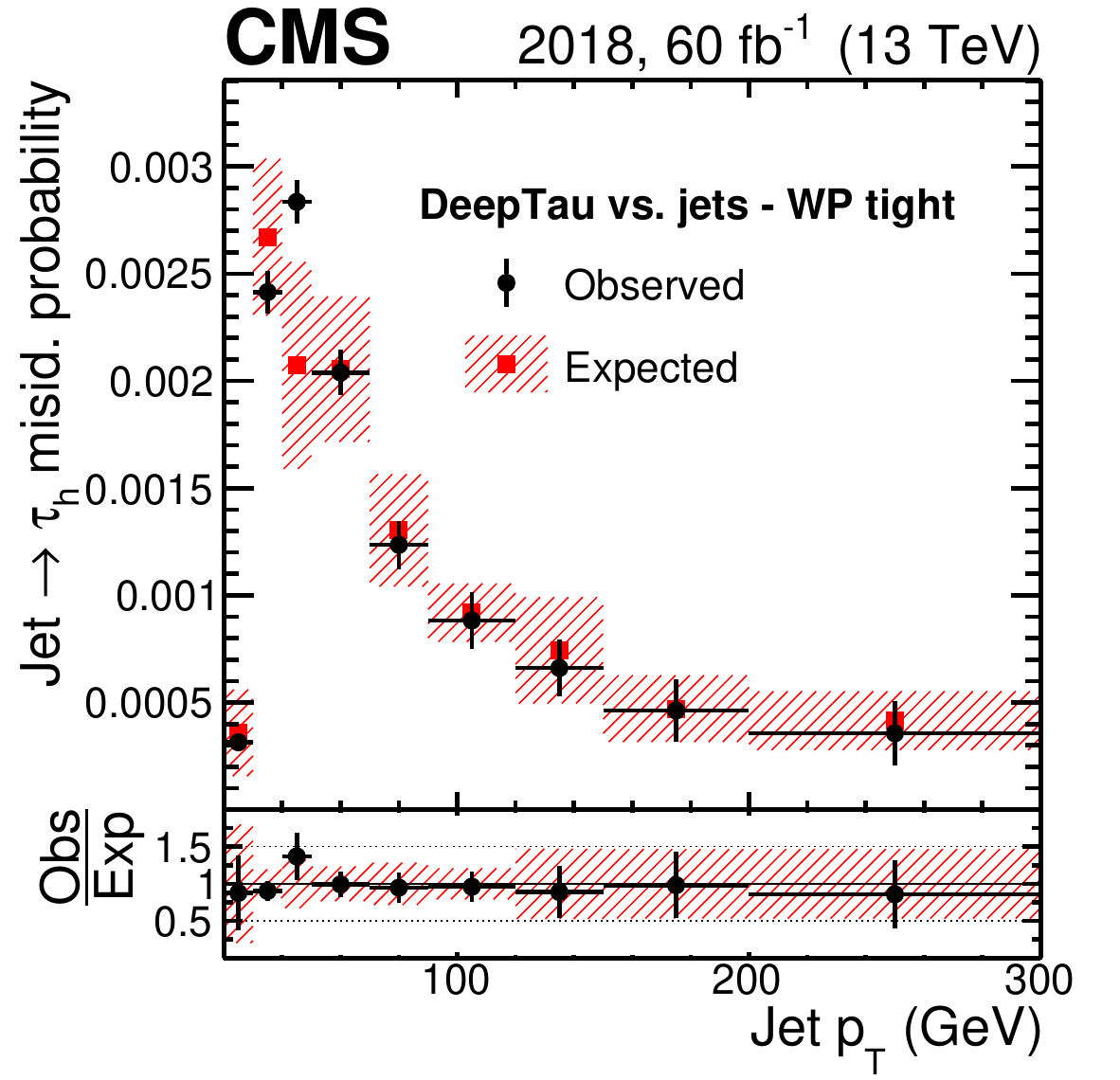}
	\caption{Observed and expected efficiencies for quark and gluon jets with $\abs{\eta} < 2.3$ to be reconstructed as \tauh candidates with $\pt > 20\GeV$ and to pass the loose (left) and tight (right) \Djet working points. The efficiencies are shown as a function of reconstructed jet \pt and are obtained with data recorded in 2018 and the corresponding simulated events.
	The shaded uncertainty band includes contributions from the limited number of simulated events and from uncertainties in the jet energy scale and the \ptmiss description.
	Besides the statistical uncertainty in the observed events, the error bars in the ratio of data to simulation also include uncertainties from the subtraction of events with genuine \tauh candidates, electrons, or muons. 
	The efficiency first rises with \pt near the 20\GeV threshold because it becomes more likely for a jet to give rise to a reconstructed \tauh candidate that passes this threshold.
	For higher \pt, the particle multiplicity in a quark or gluon jet increases with \pt. Therefore, the jets become easier to distinguish from genuine \tauh candidates and the probability for quark or gluon jets to pass the \Djet discriminator decreases with \pt.}
	\label{fig:sf_jet}
\end{figure}

For a measurement of the efficiency for jets to pass the \Djet discriminator, events are selected that are consistent with a $\PW(\to\Pgm\Pgngm)$+jet topology.
Similar to the previously discussed measurement, the events are recorded with a single-muon trigger with a \pt threshold of 27\GeV, and the reconstructed muon is required to have $\pt > 28\GeV$.
Furthermore, there must be exactly one central jet with $\pt > 20\GeV$ and $\abs{\eta} < 2.3$, and events with additional reconstructed muons or a reconstructed electron passing loose identification criteria are rejected.
To select a pure sample of $\PW$+jets events, the transverse mass of the muon and \ptmiss is required to be greater than 60\GeV.
The reconstructed jet serves as a probe for the measurement of the jet-to-\tauh misidentification efficiency.
The efficiencies for jets to be reconstructed as \tauh candidates and to pass the Loose and Tight \Djet working points are shown in Fig.~\ref{fig:sf_jet} as a function of reconstructed jet \pt.
In the measurement of the efficiency, background events with genuine \tauh candidates are subtracted.
The observed and expected efficiencies agree within the uncertainties.

\section{Summary}
\label{sec:summary}

A new algorithm has been introduced to discriminate hadronic tau lepton decays (\tauh) against jets, electrons, and muons.
The algorithm is based on a deep neural network and combines fully connected and convolutional layers.
As input, the algorithm combines information from individual reconstructed particles near the \tauh axis with information about the reconstructed \tauh candidate and other high-level variables.
In addition, an improved \tauh reconstruction algorithm is introduced that increases the overall efficiency of the reconstruction by explicitly considering the \tauh decay mode with three charged hadrons and a neutral pion, and by applying looser quality criteria for the charged hadrons in the case of three-prong \tauh decays.
The performance of the new \tauh identification and reconstruction algorithms significantly improves over the previously used algorithms, in particular in terms of discrimination against the background from jets and electrons.
For a given jet rejection level, the efficiency for genuine \tauh candidates increases by 10--30\%.
Similarly, the efficiency for genuine \tauh candidates to pass the discriminator against electrons increases by 14\% for the loosest working point that is employed in many analyses.
Following its superior performance, CMS physics analyses with tau leptons will significantly increase their sensitivities when using the new algorithm.
The superior performance of the algorithm is validated with collision data.
The observed efficiencies for genuine \tauh, jets, and electrons to be identified as \tauh typically agree within 10\% with the expected efficiencies from simulated events.
The agreement is similar to the one observed with previous algorithms and confirms the improvements.

\begin{acknowledgments}
	We congratulate our colleagues in the CERN accelerator departments for the excellent performance of the LHC and thank the technical and administrative staffs at CERN and at other CMS institutes for their contributions to the success of the CMS effort. In addition, we gratefully acknowledge the computing  centers and personnel of the Worldwide LHC Computing Grid and other  centers for delivering so effectively the computing infrastructure essential to our analyses. Finally, we acknowledge the enduring support for the construction and operation of the LHC, the CMS detector, and the supporting computing infrastructure provided by the following funding agencies: BMBWF and FWF (Austria); FNRS and FWO (Belgium); CNPq, CAPES, FAPERJ, FAPERGS, and FAPESP (Brazil); MES and BNSF (Bulgaria); CERN; CAS, MoST, and NSFC (China); MINCIENCIAS (Colombia); MSES and CSF (Croatia); RIF (Cyprus); SENESCYT (Ecuador); MoER, ERC PUT and ERDF (Estonia); Academy of Finland, MEC, and HIP (Finland); CEA and CNRS/IN2P3 (France); BMBF, DFG, and HGF (Germany); GSRI (Greece); NKFIA (Hungary); DAE and DST (India); IPM (Iran); SFI (Ireland); INFN (Italy); MSIP and NRF (Republic of Korea); MES (Latvia); LAS (Lithuania); MOE and UM (Malaysia); BUAP, CINVESTAV, CONACYT, LNS, SEP, and UASLP-FAI (Mexico); MOS (Montenegro); MBIE (New Zealand); PAEC (Pakistan); MSHE and NSC (Poland); FCT (Portugal); JINR (Dubna); MON, RosAtom, RAS, RFBR, and NRC KI (Russia); MESTD (Serbia); MCIN/AEI and PCTI (Spain); MOSTR (Sri Lanka); Swiss Funding Agencies (Switzerland); MST (Taipei); ThEPCenter, IPST, STAR, and NSTDA (Thailand); TUBITAK and TAEK (Turkey); NASU (Ukraine); STFC (United Kingdom); DOE and NSF (USA).

	\hyphenation{Rachada-pisek} Individuals have received support from the Marie-Curie program and the European Research Council and Horizon 2020 Grant, contract Nos.\ 675440, 724704, 752730, 758316, 765710, 824093, 884104, and COST Action CA16108 (European Union); the Leventis Foundation; the Alfred P.\ Sloan Foundation; the Alexander von Humboldt Foundation; the Belgian Federal Science Policy Office; the Fonds pour la Formation \`a la Recherche dans l'Industrie et dans l'Agriculture (FRIA-Belgium); the Agentschap voor Innovatie door Wetenschap en Technologie (IWT-Belgium); the F.R.S.-FNRS and FWO (Belgium) under the ``Excellence of Science -- EOS" -- be.h project n.\ 30820817; the Beijing Municipal Science \& Technology Commission, No. Z191100007219010; the Ministry of Education, Youth and Sports (MEYS) of the Czech Republic; the Deutsche Forschungsgemeinschaft (DFG), under Germany's Excellence Strategy -- EXC 2121 ``Quantum Universe" -- 390833306, and under project number 400140256 - GRK2497; the Lend\"ulet (``Momentum") Program and the J\'anos Bolyai Research Scholarship of the Hungarian Academy of Sciences, the New National Excellence Program \'UNKP, the NKFIA research grants 123842, 123959, 124845, 124850, 125105, 128713, 128786, and 129058 (Hungary); the Council of Science and Industrial Research, India; the Latvian Council of Science; the Ministry of Science and Higher Education and the National Science Center, contracts Opus 2014/15/B/ST2/03998 and 2015/19/B/ST2/02861 (Poland); the Funda\c{c}\~ao para a Ci\^encia e a Tecnologia, grant CEECIND/01334/2018 (Portugal); the National Priorities Research Program by Qatar National Research Fund; the Ministry of Science and Higher Education, projects no. 0723-2020-0041 and no. FSWW-2020-0008, and the Russian Foundation for Basic Research, project No.19-42-703014 (Russia); MCIN/AEI/10.13039/501100011033, ERDF ``a way of making Europe", and the Programa Estatal de Fomento de la Investigaci{\'o}n Cient{\'i}fica y T{\'e}cnica de Excelencia Mar\'{\i}a de Maeztu, grant MDM-2017-0765 and Programa Severo Ochoa del Principado de Asturias (Spain); the Stavros Niarchos Foundation (Greece); the Rachadapisek Sompot Fund for Postdoctoral Fellowship, Chulalongkorn University and the Chulalongkorn Academic into Its 2nd Century Project Advancement Project (Thailand); the Kavli Foundation; the Nvidia Corporation; the SuperMicro Corporation; the Welch Foundation, contract C-1845; and the Weston Havens Foundation (USA).
\end{acknowledgments}

\bibliography{auto_generated}
\clearpage
\appendix
\numberwithin{table}{section}
\numberwithin{equation}{section}
\section{Loss function}
\label{app:loss}

The loss function used for the training of the \textsc{DeepTau} algorithm has the form
\begin{linenomath}
\begin{equation}
  \begin{split}
    &L(\mathbf{y}^{\text{true}}, \mathbf{y}; \boldsymbol{\kappa}, \boldsymbol{
    \gamma}, \boldsymbol{\omega}) 
    =\underbrace{\vphantom{Bigl(\kappa_{\Pe}+\kappa_{\Pgm}+\kappa_{\text{jet}}
    \Bigr)}\kappa_{\Pgt}\,H_{\Pgt}(\mathbf{y}^{\text{true}},\mathbf{y};\boldsymbol{
    \omega})} + \underbrace{\Bigl(\kappa_{\Pe}+\kappa_{\Pgm}+\kappa_{\text{jet}}
    \Bigr)\overline{F}_{\text{cmb}}(1-y_{\Pgt}^{\text{true}},1-y_{\Pgt};
    \gamma_{\text{cmb}})} \\
    &\hspace{3.1cm}\text{(a) Separation of all }\alpha
    \hspace{1.2cm}\text{(b) Focused separation of} \\
    &\hspace{8.2cm}\Pe,\,\Pgm,\,\text{jet from }\tauh \\
    &\\
    &\hphantom{L(\mathbf{y}^{\text{true}}, \mathbf{y}; \mathbf{\kappa}, \mathbf{
    \gamma},\mathbf{\omega})=} + \underbrace{\kappa_{F}\sum\limits_{i\in\{\Pe,
    \,\Pgm,\,\text{jet}\}}\kappa_{i}\,\hat{\theta}(y_{\Pgt}-0.1)\,\overline{F}_{i}(y_{i}^{\text{true}},y_{i};\gamma_{i}).} \\
    &\hspace{3.5cm}\text{(c) Focused separation of }\tauh\text{ from }\Pe,\,\Pgm, \\
    &\hspace{3.5cm}\text{jet for }y_{\Pgt}>0.1\\
    &\\
    &\textbf{Categorical cross entropy:}\\
    &\\
    &H(\mathbf{y}^{\text{true}},\mathbf{y};\boldsymbol{\omega}) = -\sum\limits_{
    \alpha}\omega_{\alpha}\,y_{\alpha}^{\text{true}}\,y_{\alpha}\,; \qquad 
    \alpha\in\{\tau,\,\Pe,\,\Pgm,\,\text{jet}\}.\\
    &\\
    &\textbf{Focal loss function:}\\
    &\\
    &F(y^{\text{true}},y;\gamma)= -y^{\text{true}}\,\left(1-y\right)^{\gamma}
    \log(y)\,\qquad \overline{F}(y^{\text{true}},y;\gamma)= \mathcal{N}\,F
    (y^{\text{true}},y;\gamma)\,, \\
    &\\
  \end{split}
\label{eq:loss-function}
\end{equation}
\end{linenomath}
where bold fonts indicate groups of parameters; 
$H(\cdot,\cdot;\cdot)$ corresponds to the categorical cross entropy and $F(\cdot,\cdot;\cdot)$ to the focal loss function~\cite{focal_loss}; 
$\mathcal{N}$ is a factor to normalize $F(\cdot,\cdot;\cdot)$ to unity in the interval 0--1;
and $\hat{\theta}(\cdot)$ is a smoothened step function that approaches 1 for $y_{\Pgt}>0.1$. 
The focal loss terms in Eq.~\ref{eq:loss-function}\,(b) and (c) put more emphasis on the part of parameter space in which it is more difficult to separate \tauh candidates from \Pe, \Pgm, and jets.
The step function in Eq.~\ref{eq:loss-function}\,(c) disregards the part of parameter space in which the probability for the \tauh candidate to correspond to a genuine \tauh is low, for which we are not interested in optimal separation between \Pe, \Pgm, and jets.

The values and meanings of the parameters $\boldsymbol{\kappa}$, $\boldsymbol{\gamma}$, and $\boldsymbol{\omega}$ are given in Table~\ref{tab:loss-params}.
The values of the $\kappa$-factors are chosen such that the numerical values of the different components of the loss function are close to each other, while putting more emphasis on the discrimination against jets, since jets are the dominant source of background.

\begin{table}[hbt]
  \caption{
    Parameters used in the definition of the loss function for the 
    training of the \textsc{DeepTau} algorithm.
  }
  \label{tab:loss-params}
  \vspace{0.5cm}
  \centering
  \begin{tabular}{lcccclll}
    & & \multicolumn{3}{c}{Location in Eq.~\ref{eq:loss-function}} & & & \\
    Parameter & Value & (a) & (b) & (c) & \multicolumn{3}{l}{Meaning} \\
    \hline
    $\kappa_{\Pgt}$       & $2.0$ & $\checkmark$ & & & \multirow{4}{*}{Emphasis on} & \tauh & \multirow{4}{*}{separation} \\
    $\kappa_{\Pe}$   & $0.4$ & $\checkmark$ & $\checkmark$ & $\checkmark$ & & \Pe & \\
    $\kappa_{\Pgm}$        & $1.0$ & $\checkmark$ & $\checkmark$ & $\checkmark$ & & \Pgm & \\
    $\kappa_{\text{jet}}$ & $0.6$ & $\checkmark$ & $\checkmark$ & $\checkmark$ & & $\text{jet}$ & \\    
    $\kappa_{F}$          & $5.0$ & & & $\checkmark$ & \multicolumn{3}{l}{Emphasis on high \tauh efficiency} \\  [\cmsTabSkip]

    $\gamma_{\Pe}$   & $2.0$ & & & $\checkmark$ & \multirow{4}{*}{Focus on} & \Pe & \multirow{4}{*}{separation} \\      
    $\gamma_{\Pgm}$        & $2.0$ & & & $\checkmark$ & & \Pgm & \\      
    $\gamma_{\text{jet}}$ & $0.5$ & & & $\checkmark$ & & $\text{jet}$ & \\      
    $\gamma_{\text{cmb}}$ & $2.0$ & & $\checkmark$ & & & $\Pe,\,\Pgm,\,\text{jet}$ & \\ [\cmsTabSkip]

    $\omega_{\alpha}$      & varying & $\checkmark$ & & & \multicolumn{3}{l}{Sample normalization} \\
    \end{tabular}
\end{table}
\cleardoublepage \section{The CMS Collaboration \label{app:collab}}\begin{sloppypar}\hyphenpenalty=5000\widowpenalty=500\clubpenalty=5000\cmsinstitute{Yerevan~Physics~Institute, Yerevan, Armenia}
A.~Tumasyan
\cmsinstitute{Institut~f\"{u}r~Hochenergiephysik, Vienna, Austria}
W.~Adam\cmsorcid{0000-0001-9099-4341}, J.W.~Andrejkovic, T.~Bergauer\cmsorcid{0000-0002-5786-0293}, S.~Chatterjee\cmsorcid{0000-0003-2660-0349}, M.~Dragicevic\cmsorcid{0000-0003-1967-6783}, A.~Escalante~Del~Valle\cmsorcid{0000-0002-9702-6359}, R.~Fr\"{u}hwirth\cmsAuthorMark{1}, M.~Jeitler\cmsAuthorMark{1}\cmsorcid{0000-0002-5141-9560}, N.~Krammer, L.~Lechner\cmsorcid{0000-0002-3065-1141}, D.~Liko, I.~Mikulec, P.~Paulitsch, F.M.~Pitters, J.~Schieck\cmsAuthorMark{1}\cmsorcid{0000-0002-1058-8093}, R.~Sch\"{o}fbeck\cmsorcid{0000-0002-2332-8784}, D.~Schwarz, S.~Templ\cmsorcid{0000-0003-3137-5692}, W.~Waltenberger\cmsorcid{0000-0002-6215-7228}, C.-E.~Wulz\cmsAuthorMark{1}\cmsorcid{0000-0001-9226-5812}
% ...
\vskip\cmsinstskip
% ...
1:~Also at TU Wien, Wien, Austria\\
% ...
\end{sloppypar}
%%% END EDITABLE REGION %%%
% skeleton_end
\end{document}